# Broad-based experimental evidence for a hidden phase of cuprate coexistence with far-reaching implications

K.E. Gray

446 High Country Drive

Port Angeles, WA 98362

archalot063@gmail.com

ABSTRACT

Analyses of experimental data in the literature show thresholds that directly imply a coexistent superconductive (SC) and pseudogap (PG) phase that, to our knowledge, has not been previously identified. The data used emphasize the essences of d-wave cuprate SC, i.e., the superfluid density and the momentum dependence. For severe underdoping, these data imply that SC condenses conventionally onto the unaltered low-energy density-of-states of a high-temperature PG. Beyond a doping threshold, the high-temperature normal state is replaced, below the SC transition temperature, by a single energetically-favorable thermodynamic phase that embodies a more complex interplay of these orders. This previously hidden phase occupies a central portion of the SC dome, enhances the superfluid density by a factor of two or more to benefit applications, can resolve thirteen mysteries in the literature and suggests an alternative approach to cuprate SC.

1. Introduction.

The nature of cuprate superconductivity (SC) versus hole doping, p, is addressed, where p is the number of mobile doped holes per $CuO_2$ formula unit. The parent compound, at p = 0 is an insulating antiferromagnetic, core-spin state. For $0.19 < p < 0.27$, a purely SC phase of Cooper pairs condenses conventionally onto its normal-phase density-of-states N(E) at energy E. That N(E) exists above the SC transition temperature $T_c$, and is little affected by SC, as, e.g., niobium. It is widely agreed that long-range SC is found within a parabolic $T_c$ dome for $0.05 < p < 0.27$. Above $T_c$ for $0.05 < p < 0.19$, the normal state is commonly referred to as a pseudogap (PG), which is ill-defined and may include spin-, charge- and pair-density waves in ways that are not universal.

We greatly simplify that complication by focusing on SC below $T_c$ which exhibits a d-wave symmetry with respect to momentum k. Without a new theory, our strategy analyzes broad-based *experimental* data from the literature that emphasize the essences of cuprate SC. Cooper pairs are probed by superfluid density (SFD) and superconductor-insulator-superconductor (SIS) Josephson tunneling, but those techniques are insensitive to k. That void is filled by angle-resolved photoemission (ARPES) which probes k but it is insensitive to the SFD of Cooper pairs. While SIS *spectroscopy* is insensitive to both, it provides added insights. Our objective is reminiscent of the parable of an elephant and the blind men [1]. As such, a proper description of coexistence must find *consistent agreement* among a preponderance of each technique's data.

Our main result is the identification of thresholds, one versus doping p and another versus temperature T, that are found consistently in SFD, ARPES and SIS data. The substantially different behaviors across these thresholds are *prima facie* evidence for an energetically-favorable thermodynamic phase that was previously unidentified, and is thus a hidden phase. That hidden phase occupies a large central portion of the SC $T_c$-dome and is supported by numerous corroborations within and across cuprate materials and among these three techniques.

The radically different hidden phase implies these thresholds should be 1[st]-order. However, any 1[st]-order jumps may be masked by inevitable doping disorder in the local p, and thus $T_c$(p), of real samples. A disorder model is developed in Sec. 3.4 that is consistent with a 1[st]-order jump with local p values varying by about ± 0.01 around their average. Beyond this, Appendix A suggests that doping disorder in cuprates may be, otherwise, relatively benign.

The hidden phase doubles the SFD, can resolve 13 mysteries in the literature (see list in Appendix B) and provides an alternative approach to cuprate SC.

Section 2 presents the unvarnished evidence of thresholds straight from the relevant published papers including five directly copied figures. Section 3 analyzes these data within a framework of two exclusive models. We find for doping below a p-threshold, SC condenses on the normal-state PG that is present above $T_c$, and for doping above that threshold the normal-state PG is *replaced* by a hidden phase. Evidence for each model is presented in Secs. 3.1 and 3.2. While our analyses do not find a detailed microscopic picture of the hidden phase, Sec. 4 summarizes constraints it places on any hidden phase model and Section 5 compares those constraints with theories in the literature. Section 6 summarizes the impact and implications of the hidden phase.

## 2. Experimental evidence of thresholds.

Doping *thresholds* at p ≈ 0.076 - 0.1 are seen in three data sets each from ARPES, SFD and SIS spectroscopies. Companion T-thresholds, near $T_c$, are seen in one SFD and two ARPES data sets.

Seminal ARPES studies for T < $T_c$ [2,3] and T > $T_c$ [2,3,4] are important. In [2], crystals over a wide doping range, all with the $Bi_2Sr_2CaCu_2O_{8+d}$ (Bi2212) structure, were studied at low-T with p derived from $T_c$ using the empirical formula, $T_c = T_{cmax} (1-82.6 (p - 0.16)^2)$ of [5] with $T_{cmax}$ = 96 K. Peaks in their energy-distribution curves provide the gaps shown in Fig. 1, which is copied from [2]. For p ≥ 0.076, they show an expected d-wave node at momentum K = 0.5|cos ($k_x$)-cos($k_y$)| = 0, but lack a node below that p-threshold. For all p, the gaps extrapolate to a maximum value at the antinodal (AN) momentum (K = 1). The gap slopes vs K at near-nodal (NN) momenta trisect the SC dome into Regions A, B and C in Fig. 1d. The B to C threshold near p = 0.19 is not surprising, due to the loss of PG order for p > 0.19. The threshold from A to B was unexplained.

Figure 2 is from a seminal ARPES study [3] of $Bi_2Sr_2CuO_6$ (Bi2201). Above $T_c$, it shows the $\Delta_{AN}(T)$ are due to the AN PG, but they decrease below $T_c$ (arrows) to values that are too large for SC alone. That T-threshold implies both the PG and SC effect the low-T $\Delta_{AN}(T)$. The fairly constant NN slopes found in Region B of Bi2212 (Fig. 1d) agree with Fig. 3 of [3], as their NN slope is 5% larger in UD23K than in OP35K, in spite of a reduced $T_c$.

Another seminal ARPES study [4] probes the electronic structure for T > $T_c$. In Fig. 3, those results are combined with ARPES data for T < $T_c$ [2,3] and those that identify the PG scale T* at T > $T_c$ for p < 0.19 [6]. For p > 0.19, [4] found Fermi liquid behavior at all T. However, their 250 K data for p = 0.16 - 0.186 show a strange metal with incoherent AN quasiparticles. Upon cooling the sample with p = 0.186, coherent quasiparticles begin to emerge from the high-T strange metal at a T-threshold of ≈ 110 K; these features sharpen considerably and display a clear SC dip-hump feature [7] below $T_c$ = 86 K. While that sharpening may be a disorder-broadened 1st-order threshold for hidden-phase nucleation at $T_c$, it may be difficult to completely disentangle it from the PG appearing below T* ≈ 100 K for p = 0.186 [6]. Either case is consistent with a threshold for coherent-quasiparticle states at the Fermi level occurring near $T_c$.

Cooper pairs are the quintessential aspect of SC below $T_c$ and are probed by SFD, $n_s$, that is usually reported by the screening of magnetic fields, i.e., $n_s = (mc^2/4\pi e^2)/\lambda^2$, where $\lambda$ is the magnetic-field penetration depth. Seminal data in Fig. 4 are directly copied from [8] and they cover a doping range of 0.055 ≤ p ≤ 0.12 in high-quality, epitaxial films of Bi2212. Narrow x-ray rocking curves (0.3° to 0.4°) indicate *phase-pure* epitaxial films, while peaks in the real part of the sheet conductance, seen near $T_c$ in Fig. 4, indicate transition widths of ≈ 2 K. Downturns near $T_c$ were clearly identified in [8] to occur above a threshold at $T_c$ ≈ 50 K, i.e., p ≈ 0.08.

In [8], downturns were ascribed to 2D melting theory in [9] (dashed KTB lines), but the authors' later analysis [10] seemed to downplay that; those data, alone, confirm the downturn threshold at p ≈ 0.08. Proposed fluctuation explanations of SFD downturns, are discussed in Appendix C and each of them, including KTB, are shown to be problematic.

The sharp onset of $n_s(T)$ at $T_c$ in Fig. 4 is most likely a percolation threshold for long-range SC occurring near the median p of each film's macroscopically disordered doping profile.

In Fig. 5, their $n_s(0)$ are plotted vs $T_c$ to reveal a previously unrecognized companion threshold for doubling of $n_s(0)/T_c$; it occurs likewise near $T_c$ = 50 K and $p \approx 0.08$. Solid squares denote $n_s(0)$ for more systematic $n_s(T)$ data in Fig. 4 that are parallel like the rails of curved train tracks.

Break-junction SIS tunneling is well-established by cuprate studies [11-14] along with many other exotic SC and complex materials using SIS and/or Au point contacts. Note that the SIS spectra convolute N[E] from each side of the tunnel barrier so SIS peaks appear at twice the peaks in N[E]. These data from [14] show spectra for two severely-underdoped Bi2212 epitaxial films in which the top and bottom spectra were fit to a d-wave gap $\Delta(k)$ with a k-independent scattering rate $\Gamma$. A smaller gap peak is also seen for $T_c$ = 5 K (p = 0.053), implying a threshold between one and two gaps occurring somewhat below p = 0.085.

Other SIS spectra [14] are shown in Fig. 7; they range from p = 0.053 (UD 5K) to 0.23 (OD 57K). Conductance dips at V = $(2\Delta + \Omega)/e$ are found above a p-threshold at 0.105 ($T_c$ = 72 K). These have been definitively related to SC in [7], as $\Omega(p) \propto T_c(p)$ and the resonance spin excitation [15].

In Fig. 8, the low-T SIS gaps, $\Delta_{SIS}$, of [14] match low-T STM [16] and ARPES AN-gaps [2] for p > 0.1. To estimate the SC gap $\Delta_{SC}$ (short-dashed line), we interpolate between pure SC states for p > 0.19 and $\Delta_{SC}$ = 0 at p = 0.05 by assuming for $\Delta_{SC}(p)$ follows the same $T_c$-parabola. The abrupt jump of $\Delta_{SIS}$ at $p \approx 0.1$ establishes a threshold that mimics the loss of the $\Omega$ dip-hump feature in Fig. 7. For reference, the AN ARPES gaps above $T_c$ at 100 K also are shown as the long-dashed line.

Table I. Summary of experimental thresholds.

The doping p and temperature T thresholds described in Sec. 2 include a brief description, method used, cuprate material and a reference to their source, e.g., figures in this paper.

| Description | Method | Material | Reference | p-threshold | T-threshold |
|---|---|---|---|---|---|
| Region A to B NN gaps | ARPES | Bi2212 | Fig. 1d [2] | <0.076 | |
| Coherent AN state below $T_c$ | ARPES | Bi2212 | Fig. 2 of [4] | | ~$T_c$ |
| AN gaps below $T_c$ | ARPES | Bi2201 | Fig. 2 [3] | | ~$T_c$ |
| Downturns at high T | SFD | Bi2212 | Fig. 4 [8] | ~0.08 | |
| Doubling of $n_s(0)$ | SFD | Bi2212 | Fig. 5 [8] | ~0.08 | |
| Percolation threshold | SFD | Bi2212 | Fig. 4 [8] | | $T_c$ |
| One or two gaps | SIS | Bi2212 | Fig. 6 [14] | ~0.085 | |
| Dips at $2\Delta + \Omega$ | SIS | Bi2212 | Fig. 7 [14] | ~0.105 | |
| Abrupt jump in $\Delta_{SIS}$ | SIS | Bi2212 | Fig. 8 [14] | ~0.1 | |
| Percolation threshold | SFD/SIS | Y123 | Fig. 5 of [18] | | $T_c$ |

For these thresholds, summarized in Table 1, all p's are derived for pure Bi2212 from the empirical formula of [5] with $T_{cmax}$ = 96 K. However, there are precision limits on p for each technique; these are assessed in Appendix D. The encircled SIS gaps in Fig. 8 may be among the most reliable, since p, from $T_c$, and $\Delta_{SIS}$ are found in the *identical microscopic junction volumes*, with $T_c$ resulting from the loss of each junction's Josephson supercurrents or SC gap features.

Such reliability for p may not always be possible; e.g., probes of surface bilayers or when cation doping is needed for severe underdoping if $T_{cmax}$ is not independently determined for each cation substitution. For the SFD, the entire macroscopic volume is used for both, but due to inevitable doping disorder, $T_c$ is likely the percolation threshold for long-range SC. In summary, one cannot expect precise agreement of p-thresholds across materials and techniques.

It should also be recognized that p is a single-valued proxy for $T_c$, using the above formula [5]. There are data to support its universality and it is commonly used as such. But absolute values of $p(T_c)$ may vary with cuprate material, and p derived from APRES in Bi2212 [17] find uniformly larger p-values than [5] by 0.02.

The above thresholds provide *prima facie* evidence for a hidden phase in Region B. The T-threshold is consistent with the c-axis conductance of SIS Josephson interbilayer junctions in underdoped Y123. Those data [18] are shown in Appendix E to fit extremely well to the Ambegaokar-Halperin theory [19] for dissipation in over-damped Josephson junctions. However, the fit values of the Josephson energy $e_J(T)$, proportional to the Josephson critical current $I_{cJ}(T)$, remained linear much too far below $T_c$ for the Ambegaokar-Baratoff theory [20]. Instead, the $e_J(T)$ emulate the linear increase of $n_s(T)$ in Fig. 4 for $T_c > 50$ K and $I_{cj}(T)$ in Fig. 10. As such, both may be explained by a hidden-phase T-threshold, broadened by inevitable doping disorder (Sec. 3.4), to give the first viable explanation for the Y123 data of [18].

## 3. Analyses.

The above literature data are analyzed in two exclusive models. In Case 1, SC condenses on $N_{PG}(E)$ of the PG, which not only has a high-energy peak but $N_{PG}(E)$ must be *finite* at the Fermi level for Cooper pairs to form. A telltale signature of Case 1 is a second smaller SC gap peak appearing below $T_c$ on the otherwise unaltered $N_{PG}(E)$ found above $T_c$, as is seen in Fig. 6. Everything else comprises Case 2, in which the high-T PG phase is *replaced* below $T_c$ by an energetically-favorable phase with a more complex interplay of these orders. To avoid the tail wagging the dog in Case 2, this phase cannot nucleate if the strength of either order is too weak. For example, SC first appears at p = 0.05 and its strength increases with p. The SC strength must be a finite fraction of the PG strength for a different phase to be energetically favorable. That implies a p-threshold at p > 0.05, as is seen in Figs. 1 and 4 - 8.

The SIS data in Fig. 6, for $T_c$ = 5 K and p = 0.053 (Region A of Fig. 1c), shows that a second smaller SC gap condenses conventionally below $T_c$ on the unaltered normal-phase $N_{PG}(E)$, ala Case 1. In the absence of a PG (Region C for p > 0.19), [4] also found that SC condenses conventionally on a normal-state Fermi liquid, akin to Case 1. Region B differs greatly from Region A in that the high-T PG is replaced below $T_c$ by a single intermediate gap peak in the SIS data, see Figs. 2 and 8. This implies that the hidden phase of Case 2 is radically different from Case 1 coexistence, see, e.g., SIS data in Fig. 6 from [14] which provided that first definitive distinction between Cases 1 and 2.

### 3.1. Evidence of Case 1 in Region A.

As expected for Case 1, additional gap peaks, smaller than the PG, are seen at low-T in SIS (Fig. 6), ARPES (Fig. 1a) and, see Appendix A, in STM [21]. These small gap magnitudes are all closely proportional to $T_c(p)$, to be consistent with a purely SC gap, as expected for Case 1. The ARPES data for p = 0.063 reveal a 16 meV NN gap, that closely scales to $\Delta_{SC}(p)$ in Fig. 8, relative to the 4 meV SIS gap of Fig. 1 at p = 0.053; that is to be expected as both are Case 1. In an earlier paper [22], deeply underdoped $Bi_2Sr_2(Ca,Y)Cu_2O_8$ in Region A (p = 0.07, 0.075 and 0.08) showed AN-gaps that *increased* with less doping, ala the PG, but also smaller NN coherent quasiparticle peaks that *decreased* with less doping, ala the SC $T_c(p)$. These data suggests that the finite N(E) *at the Fermi level* needed for Case 1 in Region A may be the NN Fermi arcs that do not extend to the AN. Such k-segregation in Region A would not preclude its absence in Region B (see Sec. 3.5).

The larger SIS gap peak for data in Fig. 6 with p = 0.053 emulates the ARPES PG at 100 K [6] and the two-magnon Raman exchange constant [23], ala Case 1. Such dual gap peaks comport with Case 1 and not the single gap peaks of Region B. The lack of conductance dips past the larger SIS gap peaks in Region A (Fig. 7) implies a non-SC origin, like the unaltered PG of Case 1.

The lack of a readily visible second smaller gap in Fig. 6 with a *bulk* $T_c$ = 51 K and p = 0.085 is puzzling, vis-à-vis Fig. 8. Without a direct measure of $T_c$ in the SIS junction, that p-value might not represent a local p closer to the threshold for Region B. Then it would be more subject to the disorder broadening (see Sec. 3.4), that is readily seen in the UD70K data at p = 0.1 in Fig. 7.

Finally, downturns in SFD (Fig. 4) are absent in Region A to comport with these films being Case 1 and below the p-threshold for hidden phase nucleation of Case 2.

### 3.2. Evidence of Case 2 in Region B.

The SFD downturns of Fig 4 and relatively larger $n_s(0)/T_c$ of Fig. 5 are only seen in Region B; these imply different physics that can be reasonably understood by the hidden phase of Case 2. In Region B of Fig. 7, the ubiquitous presence of only one SIS gap with its prominent SC-related conductance dip are expected for a hybrid gap of the new thermodynamic hidden phase, ala Case 2. In Region B of Fig. 8, low-T gap peaks in SIS, STM and AN ARPES are well below the high-T PG and above the SC gap; that is also expected for a hybrid gap of a Case 2 hidden phase.

There is a T-threshold, only found in Region B, for a more *abrupt* onset of the SFD at $T_c$ (see Fig. 4), which signals a percolation threshold for the nucleation of the hidden phase of Case 2. Another T-threshold is the ARPES spectra [4] for p = 0.186, that show a strange metal above, and coherent quasiparticles below, $T_c$.

### 3.3. Superfluid density.

All SFD data in Fig. 4 can be parameterized by $n_s(T) = n_s(0)(1-(T/T_{c2})^2)$. These are shown as dashed lines in Fig. 9 for the more systematic data of Fig. 4 that are parallel like the rails of curved train tracks. For films with downturns, the low-T data, alone, define those fits.

Now consider the end points of the dashed lines in Fig. 9 at $T_{c2}$. Without a hidden-phase enhancement, SC is likely too weak to condense at $T_{c2}$. But as T decreases, SC should eventually appear as it does in the data of Fig. 9 for $T_c$ < 50 K. Then as T decreases further, SC strengthens until it passes the hidden-phase threshold and, assuming no disorder, $n_s(T)$ would be expected to jump to values appropriate to dashed lines ending at $T_{c2}$. Even slight doping disorder can easily disguise that 1$^{st}$-order jump (Sec. 3.4 below and Appendix A) and the very first appearance of SFD, as T decreases, would be unlikely to percolate SC throughout the films.

### 3.4. Doping disorder.

Doping disorder may exist over various length scales for individual samples. Even with a uniform average p over macroscopic scales, disorder in local p over nanoscopic scales seems inevitable, due to the low hole density in highly underdoped cuprates. It has been seen by STM [21,24], but samples may also exhibit mesoscopic disorder. Most data used here can be understood by a simple model of mesoscopic and nanoscopic disorder, but some data (samples) imply macroscopic disorder. The less systematic data in Fig. 9, without dashed lines, appear to show some *macroscopic* doping disorder. Such data might be disregarded.

The scale of disorder can be identified in some epitaxial Bi2212 films. In Fig. 10, the SFD data *below* the downturn represent the hidden phase filling the entire film area; and as in Fig. 9, they can be parametrized by the upper dashed line with $T_{c2}$ = 97 K. The Josephson supercurrents $I_{cJ}(T)$ in an SIS junction on a seperate film with the same $T_c$ comports to the Ambegaokar-Baratoff theory [20] with the same $T_{c2}$ = 97 K. Both these downturns are broadened the same to imply $I_{cJ}(T)$ is a proxy for $n_s(T)$ *and* that the relevant disorder is nanoscale, since broadening is similar for mm-sized SFD films and submicron-sized SIS junctions. Note, the lower dashed line in Fig. 10 is the hypothetical $n_s(T)$ for $T_c$ = 70 K *without a hidden phase*, as in the data for $T_c$ < 50 K in Fig. 9.

Virtually all our relevant data are consistent among techniques with such mesoscopic disorder showing a p-width of $\delta p \approx \pm 0.01$ about its nominal value. Most SFD data in Figs. 4 and 9 recover values appropriate to $T_{c2}$ (with the hidden phase filling the entire film area) at $T \approx T_c$ - 10 K. For the $T_c$ = 70 K data in Fig. 10, that T-difference implies p varies by ± 0.01 from its 2D percolation value of p = 0.102. Another example of $\delta p \approx \pm 0.01$ is the abrupt jump of $\Delta_{SIS}$ near p = 0.1 in Fig. 8, since data at p $\approx$ 0.085 and 0.11 follow the trends of samples further away from p = 0.1.

Note that while $n_s(0)$ is doubled with the hidden phase in Fig. 5, $n_s(T)$ is up to five times larger near $T_c$ in Fig. 10, an enhancement that is only limited by $\delta p = \pm 0.01$; if $\delta p$ were zero, there would be no broadening but rather a 1$^{st}$-order vertical jump in $n_s(T)$. For films with $T_c$ = 69 to 79 K in Fig. 4, $\delta p$ varies from ± 0.008 to ± 0.012.

Such a mesoscopic disorder model can explain data in Fig. 4 for $T_c \approx$ 54 K which exhibits downturns but its $n_s(0)/T_c$ in Fig. 5 is far below the systematic behavior of the larger-$T_c$ films with downturns. These films are near the p-threshold between Regions A and B, so doping disorder can lead to a mixture of local patches with, and without, the hidden phase. To test this, the diamonds in Fig. 9 are a simple sum of 50% of extrapolations to $T_c$ = 54 K of both the systematic high- and low-$T_c$ data; the modest agreement reinforces such a mixed-phase model.

Finally, doping disorder from STM gap maps in severely underdoped Bi2212 [21,24] is reviewed in Appendix A and it seems to be of minor consequence beyond the above discussion.

### 3.5. Momentum segregation of SC and PG.

A persistent theme in the literature is segregation of the SC and PG orders in momentum space. However, the ARPES data in Figs. 1 and 2 run counter to that argument. The authors of [2] recognized that the nearly-constant, near-nodal (NN) gap slopes in Region B meant that NN states cannot be purely SC, since the superfluid density of Bi2212 increases linearly across the entire Region B, starting from zero at p = 0.05 [25]. This behavior in Region B of Fig. 1d is also seen in

single-CuO$_2$-layer cuprates Bi$_2$Sr$_2$CuO$_{6+d}$ (Bi2201), where the NN slopes were quantified by $\Delta_o$ in Fig. 3 of [3]; they find $\Delta_o$ = 18 meV for OP35K but a bit larger (19 meV) for underdoped UD23K in spite of the reduced T$_c$ (35 K to 23 K). Then the NN gaps cannot be predominantly or wholly SC, but are *affected by both SC and the PG* in Region B, at least for underdoping, p ≲ 0.16.

Below T$_c$, the authors of [2] also recognized that SC suppressed the AN-gaps to values less than the pure PG above T$_c$, at least for p < 0.16 in Region B in Fig. 8. For p > 0.16, this trend is directly seen in Fig. 2 for $\Delta_{AN}(T)$ in optimum and overdoped Bi2201. Except for OD18K (possibly in Region C), the Fig. 2 data from [3] shows $\Delta_{PG}(T)$ well above T$_c$, with a reduced hybrid gap below T$_c$ (arrows). Thus, both [2] and [3] show that low-T AN gaps are not predominantly or wholly PG, but are *affected by both SC and the PG* in Region B.

If NN and AN gaps are substantially affected by both orders in Region B, we infer the same for intermediate k-states. Further, single gap peaks in k-dependent (Fig. 1) and k-integrated (Fig. 8), spectra are seen ubiquitously throughout Region B, and they could not reasonably result from k-*segregated* PG and SC gaps. If all k-states manifest both, *neither order occupies its own sector of k-space*, and both orders pervade all k-states in Region B. Importantly, these data reinforce that the hidden phase in Region B must be a single thermodynamic phase embodying all k states.

### 4. Constraints on the hidden phase.

Based on experimental p- and T-threshold data, our analysis concludes that the PG is replaced below T$_c$ in Region B by an energetically-favorable phase, which was previously hidden from view. Those data dictate constraints on the nature of that hidden phase that are summarized here.

Figure 8 shows the same *single* gap peaks by ARPES [2], SIS [14] and STM [16] throughout Region B (and Region C), while in Region A, a second smaller SC gap appears below T$_c$ on the otherwise pure PG N(E) seen above T$_c$ [2,14,22]. Figure 8 also shows the single gaps in the *underdoped* portion of Region B are significantly reduced from the pure PG phase measured above T$_c$ [6], and significantly above $\Delta_{SC}$ estimated from $\Delta_{SC} \propto$ T$_c$, to imply a *hybrid* excitation gap. That same behavior is shown in Fig. 2 for optimum and overdoped Bi2201 crystals in Region B. Figure 4 shows high-T downturns in the SFD only in Region B. Figure 5 shows the low-T SFD, n$_s$(0)/T$_c$, in Region B is twice that of Region A. A direct conclusion of ARPES data [2,3,4] in Figs. 1 and 2 is that all momentum states in Region B are affected by both SC and the PG, so there is no k-segregation of SC or the PG to imply a single thermodynamic phase.

### 5. Candidate theories for the hidden phase.

Having ruled out k-space segregation, two candidate theories in the literature for the hidden phase are considered. First is the theory of coupled Ginzburg-Landau order parameters [26]; but they might be ruled out since they predict separate, unequal gaps for each of the coupled order parameters in Region B, whereas unequal gaps are absent from both k-dependent (ARPES) and k-integrated (SIS) spectroscopies. The theory of intertwined orders in nanoscale stripes [27] states no such requirement. Their nanoscale stripes, within planar CuO$_2$ bilayers, alternate between metallic stripes, hosting all mobile holes, and thus SC Cooper pairs, and a core-spin PG. Cooper pairs are assumed to Josephson couple across insulating core-spin stripes for robust SC.

How might this stripe theory [27] conform with data-dictated constraints on the hidden phase. Such a stripe phase seems unlikely to form if either SC or the PG are too weak, so it should exhibit p- and T-thresholds as found in Sec. 2. The enhanced $n_s(0)/T_c$ in a hidden phase (Fig. 5) may be explained as the effect of doubling the local p in the metallic stripes outweighing the absence of Cooper pairs in the other half of the volume.

Note that such a volume localization already occurs along the c-axis of cuprates as they exhibit insulating layers between SC layers. Another analogy occurs with nanoscale stripes [27], where the free energy is lowered by concentrating mobile holes, and SC, in one-half of the volume, i.e., their metallic stripes. Such localization also seems necessary to explain an isolated, five-nm-diameter patch, whose $\Delta_{STM} \approx 35$ meV implies near-optimal doping (p = 0.16) from Fig. 8; it was found amidst the *most underdoped* STM gap map in Fig. 2c of [24] that exhibits an average p of 0.11 (see Appendix A). Perhaps sparse mobile holes concentrate in patches to lower the free energy by condensing more robust SC there.

Nevertheless, in *isolated* spin stripes, the excitation energy would be the ARPES PG found *above* $T_c$, whereas in *isolated* metallic stripes, it would be the smaller Cooper-pair binding energy. In Region B, the single gaps are intermediate between these limits (see Fig. 8), so one needs to envision a mechanism for nanoscale stripes of [27] to exhibit a single, hybrid *excitation* gap.

There is a pathway if hole *excitations*, like Cooper pairs, are not localized in metallic stripes. Then the PG and SC gaps might hybridize into one gap, in an analogy to the dirty limit of SC with a k-dependent gap anisotropy [28]. Such a dirty limit is defined by the mean-free-path for elastic scattering being less than the SC clean-limit coherence length, $\xi_o = \hbar v_F/\pi\Delta_{SC}$, where $v_F$ is the Fermi velocity. Then excitations toggle rapidly between gaps at different momenta to exhibit a single hybrid gap that is between the values at the extremal momenta [28]. For nanoscale stripes, the equivalent of the mean-free-path is the distance traveled across the width of insulating core-spin stripes, i.e., a few lattice constants or $\approx 1$ nm. To compare this with $\xi_o$, note that the NN gap slopes in Fig. 1d for Region B, extrapolate to a *maximum* $\Delta_{SC}$ of $\approx 40$ meV *at the AN*. Using $v_F$ = 2.5x10$^7$ cm/s [29] one finds $\xi_o$ = 1.3 nm. But $\xi_o$ might be considerably larger if the smaller d-wave average of $\Delta_{SC}(k)$ is relevant. Also, for strong underdoping, $\Delta_{SC}$ will be smaller due to (i) the 40 meV AN gaps being not of a purely SC origin and (ii) that $\Delta_{SC}(p)$ mimics $T_c(p)$ in Fig. 8 and the drastic reduction of SFD at lower p [25]. Thus, a hybrid gap cannot be easily ruled out.

Such a hybrid gap provides a reasonable explanation for the reduced AN gaps below $T_c$ in Figs. 2 and 8. For example, in optimally doped Bi2201 (OP35K), the low-T SC gap *without the hidden phase* might be $\approx 15$ to 20 meV, based on data for the largest p, that is possibly in Region C and thus lacks a hidden phase. If the high-T PG in Fig. 2 for OP35K is extrapolated to low T, it would be $\approx 45$ meV. Then the measured 35 meV gap at 10 K, which is too large to be purely SC, might comport with such a hybrid gap. Also, the nearly-constant NN gap slopes vs p in Region B of Fig. 1d may be justified by hybrid gaps that are between the PG, which decreases with p [6], and the SC gap, which increases with p in concert with $T_c$ and the SFD (25).

In summary, it seems that the theory of intertwined orders in nanoscale stripes [27] might fulfill the constraints on the hidden phase including its single, hybrid excitation gap, but coupled Ginzburg-Landau order parameters [26] would not seem to do so.

6. Impact of principal findings.

The preponderance of thresholds in ARPES, SFD and SIS corroborate our principal finding of a hidden phase below $T_c$ in Region B. That hidden phase can resolve 13 previously unexplained mysteries in the literature (Appendix B). The hidden-phase threshold at $T_c$ [4,14] rules out a prominent literature theme that SC condenses on the PG existing above $T_c$, which importantly, implies the PG states above $T_c$ may be less relevant to SC in the hidden phase. The T-threshold for a hidden phase is seen in ARPES [4] up to the highest doping level (p = 0.186) of Region B. The hidden phase doubles the SFD $n_s(0)$ in Fig. 5, at least for the only available data at 0.1 < p < 0.12. Near $T_c$, $n_s(T)$ is five times larger in Fig. 10, a factor that is only limited by disorder broadening of $\delta p = \pm 0.01$. That enhancement can be greatly beneficial to applications. Notably, such $\delta p$ may plague the direct observation of the first-order nature of hidden-phase nucleation boundaries, but Appendix A suggests disorder is otherwise benign to leave our principal findings intact.

In their quest for significant terahertz (THz) radiation from the ac Josephson effect in c-axis interbilayer junctions of Bi2212 mesas, the authors of [30] were intimately familiar with the effects of heating visibly described in Appendix F for microscopic mesas (area of one $\mu m^2$ with ten c-axis junctions). Fortunately, the first two authors ignored this and their Queen-Mary-sized mesas (of area 12000 $\mu m^2$, containing 600 c-axis junctions) emitted copious THz radiation. That incongruity might be the 14$^{th}$ mystery explained by the hidden phase. Their underdoped Bi2212 mesas are in Region B ($T_c$ = 77 - 80 K and p $\approx$ 0.114) and the radiation from one mesa peaked for self-heating to $\approx$ 50 K but persisted as high as $\approx$ 60 K [31]. For $T_c$ = 80 K (70 K), Fig. 9 shows the hidden phase occupies 100% of the sample below $\approx$ 65 K (55 K), but those fractions and their unusually enhanced SFD drop precipitously above 65 K (55 K). That coincidence seems to be a concrete demonstration of the real-world benefits of the hidden phase.

We find our constraints on the hidden phase to be consistent with the theory of nanoscale stripes [27] with only very minor additions as described in Sec. 5. Our results beg to test the generality of this beneficial hidden phase in other coexistent superconductors.

For example, recent pulsed-pump-probe spectroscopy of La$_{1.6-x}$Nd$_{0.4}$Sr$_x$CuO$_4$ (LNSCO) [32,33] shows evidence of a transient charge-density wave (CDW) above $T_c$ that emulates some aspects of the stripe-phase theory [27]. Those authors suggest that it *might* imply transient SC above $T_c$, but without proof [33]. In equilibrium, these LNSCO are SC below 10 K, but exhibit low-conductivity stripe or CDW patterns slightly above $T_c$. That behavior has similarities to Bi2212 in Region B, which exhibits a low-conductivity PG state above $T_c$ and long-range SC below $T_c$. This invites the possibility that LNSCO below $T_c$ is akin to the hybrid hidden phase of Bi2212. If so, one may envision that the stripe-like CDW features above $T_c$ [33] result from a laser-driven transient extension of the hybrid hidden phase to T > $T_c$. The connections of LNSCO to the Bi2212 hidden phase are through their existential needs for another order competing with SC (PG or CDW) and their enhancement of the superfluid density (Fig. 5), even perhaps to above $T_c$. Such a conjecture would obviously need further examination, both experimentally and theoretically.

Above $T_c$, there is a much smaller inventory of ARPES and SIS data than below, but our analysis in Appendix F finds possible consistency with a modified Fig. 3. It suggests strange-metal behavior might persist above $T_c$ throughout Region B In Bi2212. Note that at p = 0, the core-spin state exists without quasiparticles, so one may surmise that the strange metal without well-defined quasiparticles could coexist with a PG. However, in Region A the states-conserving PG is clearly visible below $T_c$ by quasiparticle tunneling, see Fig. 6, but also for $T > T_c$, see Appendix F. Then these data suggest our threshold at $p \approx 0.1$ is also a lower bound for the high-T strange-metal in Bi2212 just as p = 0.19 is its upper bound. The sparser data above $T_c$ necessarily lack the corroboration across multiple techniques and cuprate materials that the hidden phase has, but to our knowledge, this provocative suggestion has not appeared explicitly in the literature. Its ultimate importance is that the normal state above $T_c$ may be less relevant to SC below $T_c$ in Region B of Bi2212. It is also difficult to visualize how well-defined SC quasiparticles in Bi2212 could evolve continuously from a strange metal, ala the $2^{nd}$-order transition at $T_c$ of conventional SC.

Somewhat tangentially, existing ARPES data [2,3] are shown here to argue *against* the SC and PG orders being momentum segregated in Region B, and thus against the PG being absent at NN momenta [34]. Also, in conjunction with SIS spectra [7,13,14], these argue against *two* gaps being ubiquitous [35]. Appendix C argues against the SFD downturns in Fig. 4 being due to fluctuations [9,36] or strong pair-breaking scattering rates [37].

## 7. Acknowledgements

The author is greatly indebted to J.F. Zasadzinski for ongoing discussions and permission to use data from their preprint [14], which was the inspiration for this research enquiry, and provided directly copied Figs. 6 and 7. Others who provided additional foundational data for our analyses and conclusions include: T.R. Lemberger for permission to directly copy Fig. 4 from [8]; Inna M. Vishik for permission to directly copy Fig. 1 from [2]; and Adam Kaminski for permission to directly copy Fig. 2 from [3]. Informative communications and advice from Mac Beasley, Jeff Tallon, John Tranquada and Mike Norman are gratefully acknowledged. Otherwise, support for this work came solely from the author.

Appendix A: Effect of doping disorder $\delta p$.

Minor disorder seems fairly benign for the gap over a majority of the doping range, but $\delta p$ must matter for the vastly different $\Delta_{SIS}$ across 1st-order transitions, e.g., the Region A to B transition in Fig. 8 at p ≈ 0.1. Section 3.4 showed consistency of SFD and SIS data with $\delta p \approx \pm 0.01$ about its nominal value. For the Region B to C transition, $\Delta_{SIS}$ varies so smoothly with p that it should be fairly insensitive to minor $\delta p$ and the rapidly diminishing PG as p approaches 0.19. These likely preclude any noticeable threshold or other dramatic effects. In Region A, SC condenses conventionally on the PG so $\Delta_{SC}$ is expected to mimic $T_c(p)$ as shown in Fig. 8. Those $\Delta_{SC} \propto T_c(p)$ would be less dependent on $\delta p$, in a similar manner to Region C.

Due to a radically different hidden phase, the thresholds at p ≈ 0.1 and T = $T_c$, are expected to be 1st-order, but that could be masked by inevitable p-disorder. Since $T_c$ depends on p, experimental data near *each* threshold could be affected. Spectra are averages over macroscopic areas for ARPES but microscopic volumes for SIS. In Region B of Fig. 8 for p > 0.1, ARPES and SIS give similar average gaps, to imply that relevant disorder is primarily sub-micron.

In Fig. 8, very different extrapolations of the average $\Delta_{SIS}$ from above and below p ≈ 0.1 are consistent with a 1st-order threshold. But, the rogue $\Delta_{SIS}$ at p = 0.1 falls midway between the extrapolated limits, suggesting a mixed phase consisting of patches with or without the hidden phase; that was corroborated by the SFD data at $T_c \approx 54$ K in Sec. 3.3. A mixed phase should have SIS peaks for the tunneling channels, A to A, A to B and B to B; these are predicted to be 210-230 mV, 152-175 mV and 105-120 mV, respectively, by extrapolating the tends in Fig. 8 to p = 0.1.

Then, can the single $\Delta_{SIS}$ peak at p = 0.102 in Fig. 7 result from such a mixed phase? Notably, *compared to all neighboring p-values*, its low-voltage conductance is clearly V-shaped, rather than U-shaped, and its FWHM divided by $\Delta_{SIS}$ is almost doubled. In Fig. 7, the single peak at 150 mV could be a dominant A to B channel with its 50-220 mV width extending to the A to A and B to B channels. Thus, the rogue $\Delta_{SIS}$ peak at p = 0.102 might be explainable, provided nanoscale disorder occurs within A and B patches to smear out three well-defined narrow peaks.

Fortunately, nanoscale disorder has been studied [21,24] extensively by scanning tunneling microscopy (STM). The sharp STM tip provides the tunneling conductance $\sigma(V) = dI/dV(V)$ within an area of order 1 $nm^2$, which depends on the tip to sample distance. Note that the significant nanoscopic spatial variation of high-voltage gap peaks, seen within 49x49 $nm^2$ areas in *heavily underdoped* Bi2212, is not seen for optimum-doped Bi2212 in [16].

The lowest density of interstitial oxygen dopants (white dots in Fig. 2 of [24]) occurs in gap maps that exhibit the largest variation of local gaps. Surprisingly, amidst the *most underdoped* gap map in Fig. 2c of [24], a five-nm-diameter patch exists with $\Delta_{STM} \approx 35$ meV, which implies near optimal doping from Fig. 8. Perhaps sparse mobile holes concentrate in patches to lower the free energy by condensing more robust SC. An analogy occurs with nanoscale stripes of [27], where the free energy is lowered by concentrating mobile holes, and SC, in one-half of the volume, i.e., their charge stripes. However, the static nature of STM gap maps requires a role for crystal defects, including the interstitial-oxygen dopant locations.

Furthermore, well-defined peaks are reported in [21,24] only for $\Delta_{STM} \leq 55$ meV, for which the equality of $\Delta_{STM}$ [16] and $\Delta_{SIS}$ in Fig. 8 imply they are only affected by underlying *nanoscale* disorder. All these gap values exist in the trend of microscopic spatial averages of $\Delta_{SIS}$ in Fig. 8 for p > 0.1, so each $\Delta_{STM}$ *could be assigned* to a local p-value, e.g., the five-nm patch discussed above. In the gap maps in Fig. 2 of [24], one expects the nm-sized STM spectral peaks to be narrower than those in micron-sized *superconductor-insulator-normal* (SIN) point-contact tunneling [38], and for p ≈ 0.11, the STM widths are indeed ≈ 80% of the SIN widths.

Patches with $\Delta_{STM}$ > 55 meV, in Figs. 6A and 6B of [21], were assigned to a zero-T PG (ala Case 1), but there are subtle low-voltage features that might suggest a second, smaller SC gap of 10 - 15 meV. Such a gap would comport with the 16-meV NN ARPES gap in Fig. 1a for p = 0.062 [2] as well as the SIS data with two gaps in Fig. 6 for p = 0.053; all these are in Region A.

For the three gap maps in Fig. 2 of [24], the number of interstitial oxygen dopant sites (white dot counts of 883, 580 and 455) are *their* most direct measure of each map's *average* p. Using two-mobile holes per white dot, one obtains $p_{ave}$ = 0.11, 0.07 and 0.06, and the latter two, in which very large gaps predominate, would clearly be in Region A. If SC gaps in Region A were roughly proportional to $T_c$, then the 10 to 15-meV gaps in Fig. 6B of [21] would be consistent with similar values of p = 0.056 to 0.06. Thus, both large and small gaps in [21,24] fulfill the expectations of Region A, with the larger being the zero-T PG that should be fairly immune to doping disorder.

While the above scenario may partially demystify such nanoscale disorder [21,24], another effect of this nanoscale disorder on our analysis may be small. Excessively large *local* gaps would only affect SC by *suppressing* thermal pair-breaking, as T approaches $T_c$. That might modestly enhance the SFD at higher T, but be *included* in the SFD data of Fig. 4 and our dashed-line fits in Fig. 9. Hence, our analysis of thresholds in Sec. 3 would seem to be little affected.

Appendix B: Mysteries in the literature explained by a hidden phase.

Here we summarize the ability of a hidden phase to resolve thirteen mysteries in the literature for which no verifiable explanations are known by us to have been offered.

Evidence from SIS spectra [14] and Josephson supercurrents [14,18]. Mystery **one**: two gaps at p = 0.053 and only one for p ≥ 0.085 (Fig. 6). *Case 1 coexistence in Region A requires a smaller SC gap condensing on the (large) PG N(E), whereas Case 2 in Region B requires a single hybrid gap.* Mystery **two**: the large SIS gaps in Region A agree with pure PG seen at 100 K in Fig. 8. *The larger SIS gap for Case 1 is the pure PG*. Mystery **three**: the SIS gap drops abruptly from Region A to B in Fig. 8. *Hidden-phase nucleation is $1^{st}$-order and the SIS gap for Case 2 in Region B is the smaller hybrid SC/PG gap.* Mystery **four**: The SC dip-hump feature at $\Omega$, that is proportional to $T_c$ in Regions B and C, is missing in Region A (Fig. 7). *The larger gap for Case 1 in Region A is the bare PG, which is independent of SC, and thus cannot show a SC dip-hump*. Mystery **five**: The abrupt decrease of Josephson supercurrent $I_{cJ}$ in SIS junctions [14] approaching $T_c$ in Fig. 10 is anomalous. *These $I_{cJ}$ perfectly match the SFD hidden-phase T-threshold for doping disorder of $\delta p \approx \pm 0.01$.* Mystery **six**: A linear T-dependence of the Josephson coupling energy $e_J(T)$ in c-axis interbilayer

SIS junctions in Y123 crystals is anomalous [18] with respect to accepted theory [20]. *Since $e_J(T)$ is proportional to $I_{cJ}$, it reflects the same abrupt hidden-phase threshold as $I_{cJ}$ in Fig. 10.*

Evidence from SFD. Mystery **seven**: The absence or presence of downturns in $n_s(T)$ was not convincingly explained [8,10]. *Downturns signal a T-threshold for hidden-phase nucleation in Region B, but do not occur for Case 1 coexistence in Region A.* Mystery **eight**: The ratio $n_s(0)/T_c$ in Region B doubles that of Region A in Fig. 5. *Hidden-phase enhancement is in Region B only.*

Evidence from ARPES [2,3,4,6]. Mystery **nine**: Trisection of the SC $T_c$ dome occurs at p = 0.076 and 0.19, for which the threshold at p = 0.19 was understood as the high-p endpoint of the PG, but no explanation of the threshold at p = 0.076 was given in [2]. *Their threshold at p = 0.076 is the low-p endpoint of hidden-phase coexistence.* Mystery **ten**: from [2] "Fig. 1B indicates that NN gaps are remarkably insensitive to $T_c$ in a broad doping range constituting region B, highlighting that NN gaps in region B do not reflect the bare superconducting order parameter", but with no explanation given. *The NN gap slopes in Region B reflect a hybrid gap interpolating between a PG that decreases with p and a SC gap, which increases with p like the SFD* [(25)]. *The AN-ARPES gaps agree with the SIS gaps (Fig. 8) with each exhibiting the hybrid SC/PG gap of the Case 2 hidden phase.* Mystery **eleven**: ". . . the pseudogap is suppressed by superconductivity at low temperatures . . ." [2] and the AN-ARPES gaps are reduced from the AN-ARPES PG measured above $T_c$ [6], both in Region B. Also, Figs. 3 and 4 of [3] for Bi2201 mimic the data of [2] for Bi2212. As such, by interpolating between these NN and AN observations these authors *could* have concluded that all k-states are affected by SC and the PG to dismiss k-space segregation. *The hidden phase is a single thermodynamic phase showing a single hybrid gap embodying both PG and SC order for all k, to virtually insure no k-space segregation.* Mystery **twelve**: from [2] "Fig. 4 A–C shows that the Fermi arc just above $T_c$ does not represent the only momenta where superconductivity emerges . . .". *A hidden phase replaces the PG so the latter's finite-length Fermi arcs above $T_c$ are irrelevant.* Mystery **thirteen**: The evolution of incoherent-quasiparticle strange metal in Region B at 250 K, evolving into sharp quasiparticle peaks below $T_c$, was unexplained in [4]. *It is congruent with a T-threshold for a hidden phase below $T_c$.*

APPENDIX C: Can fluctuations cause downturns in SFD?

The roles of fluctuations to cause the SFD downturns in Fig. 4 near $T_c$ have been expressed widely in [8,10,36,37]. However, fluctuation models might struggle to explain the *abrupt* appearance of downturns above the p-threshold, but even more troubling is that the *same p-threshold* is *seen ubiquitously below 10 K*, which cannot be caused by high-T fluctuations. Low-T thresholds include (i) the enhanced SFD $n_s(0)/T_c$ in Fig. 5; (ii) the ARPES Region A to B threshold at p ≈ 0.076 for NN slopes and spectral nodes in Fig. 1 [2]; (iii) the two- to one-gap threshold in Fig. 6 [14]; (iv) the jump in SIS gap peaks at p ≈ 0.1 in Fig. 8 [14]; and (v) the abrupt appearance of dip-hump structures in Fig. 7 [14] in SIS junctions at p > 0.11. Convergence across diverse low-T data and techniques emphasizes the veracity and importance of low-T, p-thresholds, which mimic the p-thresholds for high-T SFD downturns. Fluctuation explanations of downturns only apply near $T_c$ whereas hidden-phase nucleation *inherently applies* to all six thresholds.

The two-dimensional (2D) Kosterlitz-Thouless-Berezinskii theory [9] (KTB) was invoked by the authors of [8], but it was absent from their earlier paper on underdoped Y123 [39] and downplayed in the journal publication [10] that included some of their Bi2212 data from Fig. 4. In the textbook KTB calculation [40], a downturn occurs at $k_B T_{KTB} \approx \Phi_o^2/32\pi^2 \lambda_\perp$, where $\Phi_o$ is the flux quantum and $\lambda_\perp$ is the in-plane penetration length. Since the data in Fig. 4 result from the screening of perpendicular fields by in-plane supercurrents, the experimental $\lambda$ equals $\lambda_\perp$.

In theory [40], all $\lambda$ diverge as $\lambda_\perp(T) = \lambda_\perp(0)/\sqrt{(1-T/T_c)}$, which insures $T_{KTB}$ is always less than $T_c$. However, for $T_{KTB}$ to be observably below $T_c$, one can show that $\lambda_\perp(0)$ must be very large. Consider the data in Fig. 4 at $T_c = 60$ K for which $T_{KTB}$ would be $\approx 50$ K, so the textbook formula gives $\lambda_\perp(T_{KTB}) \approx 180$ µm. That conflicts with the measured $\lambda_\perp(0) = 0.63$ µm, which, if used with the standard formula, $1 - T_{KTB}/T_c = \{\lambda_\perp(0)/\lambda_\perp(T_{KTB})\}^2$, yields an unobservable $T_c - T_{KTB} \approx 7 \times 10^{-4}$ K. As such, the downturns of $n_s(T)$ in Fig. 4 are undeniably inconsistent with textbook KTB theory [40].

Notably, magnetic decoration images [41] in near-optimum doped Bi2212 crystals show well-defined vortices with an in-plane penetration length $\lambda_\perp \approx 1$ µm at 54 K, that is similar to the values of $\lambda(0)$ in Fig. 4. The reason $\lambda_\perp$ should be so small in *thick* cuprate crystals and films was raised by the author at an invited talk, circa 1990. It is that a large effective $\lambda_\perp$ requires very weak screening by the supercurrents surrounding vortices, and that is typically found *only in physically isolated* very thin layers (thickness d << $\lambda_{ab}$). These layers contain Pearl vortices [42] for which $\lambda_\perp = \lambda_{ab}^2/d$, where $\lambda_{ab}$ is the screening length for fields perpendicular to *much thicker* layers (d >> $\lambda_{ab}$). *Collective screening by each bilayer* in thick films or crystals of Bi2212 will be similar to a thick material with d > $\lambda_{ab}$. The anisotropy is irrelevant, as screening of c-axis fields do not require c-axis currents. Also, many insulating gaps of atomic dimensions only marginally reduce screening by the volume ratio, but even using only one quarter of the Bi2212 film's 100 nm thickness in [8], the short-dashed KTB lines in Fig. 4 would be almost at the abscissa.

Now consider other proposed causes of downturns [35,36]. A comprehensive theoretical approach [43] suggests thermal phase fluctuations could contribute to the parameterized *dashed lines* in Fig. 9, but abrupt downturns need a critical transition like the KTB theory, which is shown above to be hard-to-justify. One may also consider the Lindemann melting criterion [44] for 2D vortices. For an interbilayer repeat distance of s = 1.5 nm, melting might replicate the downturn temperatures in Fig. 4, but the crossover field for 2D melting is estimated to be $\approx 1$ T in Bi2212 [44] and that is far greater than the applied fields in the measurement of $n_s(T)$ in [8].

In summary, our natural explanation of *all the downturn behavior* in Fig. 4 by the p-threshold for hidden-phase nucleation seems unmatched by considerations of phase fluctuations, including the KTB transition, or by 2D melting [44].

Another analysis [35] suggested the SFD downturns are due to a steeply rising *pair-breaking* scattering rate, $\Gamma_{pair}$, as T increases to $T_c$. That calculation was done in "the overdoped regime near $T_c$ where the pseudogap and subsidiary charge-density wave order are absent", i.e., Region C of Fig. 1d. In [35], the specific heat of an overdoped Y123 crystal (p = 0.182) was fit using $\Gamma_{pair}(T)$ as an uncorroborated adjustable parameter and went on to calculate the SFD. It is concerning that such fits to SFD in [35] exhibit a prominent tail, extending above $T_c$ to their mean-field $T_{cmf}$, with *absolutely no hint of it* in the SFD data of Fig. 4. It is also *unclear* how a steeply rising $\Gamma_{pair}$ *near* $T_c$ might explain the five thresholds *below 10 K*, that all occur at the same p ≈ 0.1.

In [37], a single particle scattering rate $\Gamma_1$ is used in the Dynes' self-energy [45] but they add $\Gamma_{pair}$. That $\Gamma_{pair}$ appears in the self-energy of [37] as a phenomenological 'inverse pair lifetime', $\Gamma_0$, that Norman, et al [46] added to the Dynes' self-energy to account for fluctuating Cooper pairs *above $T_c$*. In the ARPES fits of [46], $\Gamma_0$ was needed *only above $T_c$* for underdoped Bi2212 (UD83K). Their fits to both OD82K and UD83K show $\Gamma_1$ increases steeply as T increases to $T_c$, which is entirely consistent with fits to Pb-Bi films in [45]. Then fits in [37] using $\Gamma_{pair}$, aka $\Gamma_0$, in Norman's self-energy seem unjustified below $T_c$ as they would conflict with the purpose of $\Gamma_0$ as well as the data fits of [46]. Thus, while [37] could be relevant to SC fluctuations *above $T_c$*, it is unlikely to correctly describe the downturns in Fig. 4 *below $T_c$*.

Note also that the inverse quasiparticle lifetime in Dynes, et al [45] is $\Gamma_1 = \Gamma_R + \Gamma_d$, where $\Gamma_R$ is the recombination rate into Cooper pairs and $\Gamma_d$ accounts for T-independent scattering by defects. In equilibrium below $T_c$, $\Gamma_R$ is twice the inverse pair lifetime since a pair is created when two quasiparticles recombine. However, detailed balance guarantees no net change in the Cooper pair density, so $\Gamma_R$ *cannot produce downturns in the SFD*. As shown in [45], these components of $\Gamma_1$ are consistent with the sharp increase of $\Gamma_1$ as $T_c$ is approached from below. That results from $\Gamma_R(T)$ decreasing exponentially with $\Delta/k_BT$ [47], so it is the dominate component of $\Gamma_1$ near $T_c$, so the T-independent $\Gamma_d$ must prevail at lower T as seen in [45, 46].

APPENDIX D: Assessing p for the Region A to B threshold.

To assess the precise doping level, p, of the Region A to B threshold, note that all experimental techniques use the universal empirical relation $T_c = T_{cmax} (1-82.6 (p - 0.16)^2)$ from [5], but each has their own limitations for obtaining $T_c$ and $T_{cmax}$. Importantly, the quoted $T_c$ for SFD, SIS and ARPES data may not be representative of the *same physical volume used for their data*. For SFD of macroscopic films [8] in Fig. 4, $T_c$ reflects the same physical volume but $T_c$ is slightly tainted by it likely being a two-dimensional percolation threshold at the midpoint of any local doping disorder, ± δp. See the discussion of Figs. 9 and 10 and Appendix A.

Perhaps the most trustworthy are those SIS break junctions which, after cleaving at 10 K, survive to high T to establish $T_c$ in the *identical sub-micron junction volume* as the gap spectra (see Fig. 4 of [14]). For the four outlined SIS gaps in Fig. 8 from [7,13,14], their $T_c$'s derive from the loss of Josephson supercurrents or spectral gap features, so that their derived p's are found in precisely the same volume as the gap. These use $T_{cmax}$ = 96 K for generic Bi2212. Sample degradation is minimized in SIS spectroscopy as all data are taken within a few hours of their low-T cleave.

Such safeguards may be difficult to achieve with probes of surface bilayers, like ARPES and STM. Those cleaves are also done in ultra-high vacuum below 10 K, but raising T to assess $T_c$ is almost never done as it risks surface degradation. In such cases, $T_c$ is commonly found from prior measurements, e.g., by magnetization, on the *bulk crystal* but not directly on the cleaved *surface bilayers* within the areas being probed.

Another problem occurs for severely underdoped Bi2212 crystals that is absent from the epitaxial films used for SFD [8] and SIS [14]. Epitaxial film growth, on compatible substrates, readily stabilizes, and verifies, a phase-pure, Bi2212 crystal structure [8,10] without cation doping. As a result, a p-independent universal value of $T_{cmax}$ = 96 K is valid in the above emprical relation. However, for severe underdoping of bulk crystals, the same Bi2212 *structure* may rely on cation substitutions [2,22]; then lacking measurements of $T_{cmax}$ for each cation or using a universal $T_{cmax}$ might lead to less-precise p-values.

Based on this discussion, any expectation of finding an exact universal p-threshold across techniques for the Region A to B boundary seems remote as it may be beyond the precision limits of such techniques, even if mild doping disorder were ignored. But even without identical p-values among SFD, SIS and ARPES, the universal identification of p-thresholds between 0.08 and 0.1 seems meaningful. Perhaps the most reliable of these is the midpoint of the A to B threshold of the SIS gap data [14] at p = 0.1 in Fig. 8.

Appendix E. Reanalysis of Y123 data in Ref. 18.

The c-axis transport in an underdoped Y123 crystal ($T_c$ = 65 K, p ≈ 0.1) was analyzed in [18] as a function of T and magnetic field using the Ambegaokar-Halperin theory [19] for dissipation in overdamped Josephson junctions; overdamping was confirmed by non-hysteretic transport data. Their data was fit by the *intrinsic* Josephson coupling energy, $e_j(T)$, and their principal conclusion was that its T-dependence did not follow the Ambegaokar-Baratoff theory [20],

$$e_j(T) = \pi\hbar\sigma_{cN}(T)\Delta(T)\tanh\{\Delta(T)/2kT\}/(2e^2s), \tag{E.1}$$

where $\Delta(T)$ is the superconductive energy gap and s is the bilayer repeat distance. At best, $e_j(T)$ is very weakly field dependent, e.g., in Bi2212 [48], $\Delta$ is unchanged up to 13 T and above $T_c$, for the Y123 crystal used in [18], $\sigma_{cN}$ is unchanged up to 7 T.

However, including a c-axis field, B, Abrikosov vortices are injected into the bilayers so the zero-field theory of [19] should be modified to:

$$\sigma_c(T, B) = \{b\,\sigma_{cN}(T) + (1-b)\,\sigma_{cqp}(T)\}\{I_0(E_j(T, B)/2k_BT)\}^2, \tag{E.2}$$

where b = 1 - $B/B_{c2}(T)$ is the fractional junction area occupied by normal-state vortex cores, $\sigma_{cN}(T)$ is the normal-state conductivity extrapolated from T > $T_c$, $\sigma_{cqp}(T)$ is the quasiparticle conductivity of the junction stack in zero field *without* Josephson coupling, $I_0$ is the modified Bessel function and the *extrinsic* Josephson coupling energy is $E_j(T,B) = A_{cc}\,e_j(T)$, and $A_{cc}$ is the average area of coherently-coupled regions. For uniform junctions in zero field, $A_{cc}$ is their geometric area, but $A_{cc}$ can be reduced by defects. The densities of field-induced vortices, $B/\Phi_0$, together with the structural defect density, described there by $B_0/\Phi_0$, were convincingly shown in [18] to dictate $A_{cc} = \Phi_0/(B+B_0)$.

Here, we correct two shortcomings of [18]. First, we show that the significant deviations of the fits for their largest $R_c$ (i.e., smallest $\sigma_c$) were due to [18] following the original theory [19] without a field (b = 0), thereby neglecting the first term in Eq. E.2. Figure E.1 shows fits to $\sigma_c(T, B)$ using

Eq. E.2 and the identical fitting procedure as [18]. Such a level of fidelity borders on being unique. However, the fit values of $e_j(T)$ and $\sigma_{cqp}(T)$ were virtually unchanged.

Finally, we try to understand their fitting parameter $R_{cN}*(T)$ that was unexplained in [18]. Its inverse is equivalent, within known geometrical factors, to $\sigma_{cqp}(T)$ in Eq. E.2. Our fit values of $\sigma_{cqp}(T)$ from Eq. E.2 are shown in Fig. E.2 to be proportional to the number density of quasiparticles, $N_{qp}(T)$, found by integrating the Fermi function times the reduced BCS density-of-states over a Fermi surface with a d-wave energy gap. That reasonable agreement implies that $R_{cN}*(T)$ is now also understood, and all fits in Fig. E.1 need only one T-dependent parameter, $e_j(T)$, plus a single scaling factor linking $\sigma_{cqp}(T)$ to $N_{qp}(T)$, which in principle, is calculable.

## Appendix F. Normal state above $T_c$ from ARPES and SIS.

Since we are aware of a much smaller inventory of ARPES and SIS data above $T_c$ than below, any resulting conclusions will not have the same weight as the highly corroborated evidence for a hidden phase. One important result is the SIS conductance for the epitaxial film with $T_c$ = 70 K in Fig. 4c of [14]. This Bi2212 film with p = 0.102 exhibits a small, very broad conductance maxima above $T_c$ that persists to 130 K in Fig. 4d of [14]. That feature must represent the normal state, and it was attributed in [14] to competing order, e.g., the PG. But its peak voltage of 50 meV is only half of the ARPES 100 K PG in Fig. 8. This broad maximum differed from the SIS conductance of a Bi2212 crystal in Region C, with $T_c$ = 56 K, which was flat and featureless above $T_c$ [49]. It also differed from the *uppermost* conductance curve in Fig. 6 for the Bi2212 film in Region A with average p $\approx$ 0.053. It is unlike the other three curves, since the zero-bias conductance is not zero. Thus, that particular junction is above $T_c$ at 2 K, which could easily result from its local p being closer to 0.05 due to doping disorder of $\delta p \approx \pm 0.01$. Instead, it shows well-defined quasiparticle tunneling into a much larger PG. Thus, a junction without the broad maximum is seen in both Regions A and C, and they both reasonably represent the normal-state above $T_c$ onto which SC condenses.

Clearly the broad maxima in [14] would represent an unusual normal state, possibly akin to the strange metal found by ARPES [4] from p = 0.16 to 0.186 in Region B at 250 K. Such broad maxima are also seen by c-axis SIS tunneling in numerous single crystal mesas [48,50-53]. Notably, we argue below that all such mesa data are plausibly in Region B. Then a strange metal might exist just above $T_c$ throughout Region B. Support for this picture is given below, using ARPES energy distribution curves (EDUs) taken about 10 K above $T_c$ [54] and at 100 K [6]. Such a picture would mean the dashed-line hidden-phase threshold at p $\approx$ 0.1 in Fig. 3 for Bi2212 is also a lower bound for strange metal behavior for T > $T_c$, just as p = 0.19 is its upper bound [4].

Heating is known to be a significant problem for large-volume mesas [55], defined by their area A and the number of c-axis interbilayer junctions N. Note that intercalation [50,55] increases the c-axis resistivity to reduce heating for any N, but even for the smallest practical mesa size, N $\approx$ 10 and A $\approx$ 1 $\mu m^2$, very noticeable heating effects occur [55]. These include telltale narrowing of the gap peaks and lower gap-peak voltages [55], but also $T_c$ from the c-axis conductivity will appear lower than the actual mesa $T_c$ due to heating.

Significant causes for excessive mesa heating include N = 50 to 70 in [51] and N = 10 and A = 33 $\mu m^2$ in [52]. Heating would cause the measured mesa $T_c$ to be below its *actual value*, which would imply p is actually closer to optimum doping and likely within Region B. Smaller heating might be inferred for an *intercalated* mesa of [50], N = 10, A = 10 $\mu m^2$, but their broad maximum at 174 K appears more like a normal-state background feature, possibly due to intercalation, as that feature is dissimilar to the small amplitude, broad conductance maxima of [14] discussed here. Minimal heating might be inferred by the mesa sizes of [53], N = 10, A < 1 $\mu m^2$, and their $T_c \approx 90$ K are clearly in Region B. In [48], minimal heating might be inferred by the mesa sizes, N = 10, A = 0.8 $\mu m^2$. Their mesa with $T_c$ = 80 K and their quoted p = 0.191 [48] would place it barely into Region C. However, the inset of Fig. 2 in [55] defines a figure-of-merit for heating as $\beta$ equal to the ratio of its full-width, half-maximum to the peak voltage. For the mesa of [48], $\beta \approx 0.28$, which is mid-way between the limits of no heating, $\beta \approx 0.52$, and severe heating, $\beta \approx 0.1$, to reveal some heating. More evidence of heating in [48] is their unusually low reported value of $T_{cmax}$ = 86.5 K. The actual p is not affected by mesa heating, but it can result in smaller *measured* $T_c$ values to imply p is further from optimum, 0.16. If so, the actual p would be in Region B. In summary, such broad maxima in crystal mesas are consistently seen in Region B but not in the sparse data of Regions A and C.

Next, consider the high-T ARPES data. For Bi2212, the EDCs at 100 K throughout Region B [6] appear to be non-state-conserving; that could be consistent with (i) a strange metal that lacks well-defined quasiparticles and (ii) the non-state-conserving conclusion of electronic specific heat data [56] in Bi2212 that covered all of Region B, i.e., 0.095 < p < 0.19. Importantly, the ARPES EDCs for Bi2212 at about 10 K above $T_c$ in Fig. 4 of [54] show quasiparticle peaks for Region C (p = 0.221 and 0.203) but lack that peak for $T_c$ in the range of UD65K to OD86K (i.e., 0.98 < p < 0.194). Those p match Region B with minor caveats about the rigor of p from ARPES.

Taken together, these data present a plausible case that Bi2212 exhibits a strange metal just above $T_c$, which lacks well-defined quasiparticles and occurs *throughout* Region B for Bi 2212. To our knowledge, that idea has not appeared in the literature.

Perceptive readers will recognize that ARPES data in Fig. 2 do not mirror the above trend for Bi2212. The quasiparticle PG peaks [3], appearing above $T_c$ in Fig. 2 from [3] are not found in Bi2212, e.g., in [54], and that inconsistently had been recognized [57]. One might surmise that a strange metal above $T_c$ throughout Region B is not universal for cuprates. Nonetheless, evidence of the hidden phase in Bi2201 is clearly shown in Fig. 2 by the emergence of a single hybrid gap below $T_c$, rather than the two gap peaks shown in Figs. 6 and 8 for the Bi2212 film with $T_c$ = 5 K which is in Region A without a hidden phase. Thus, a strange-metal normal state is not required for the hidden phase.

Figures with Captions.

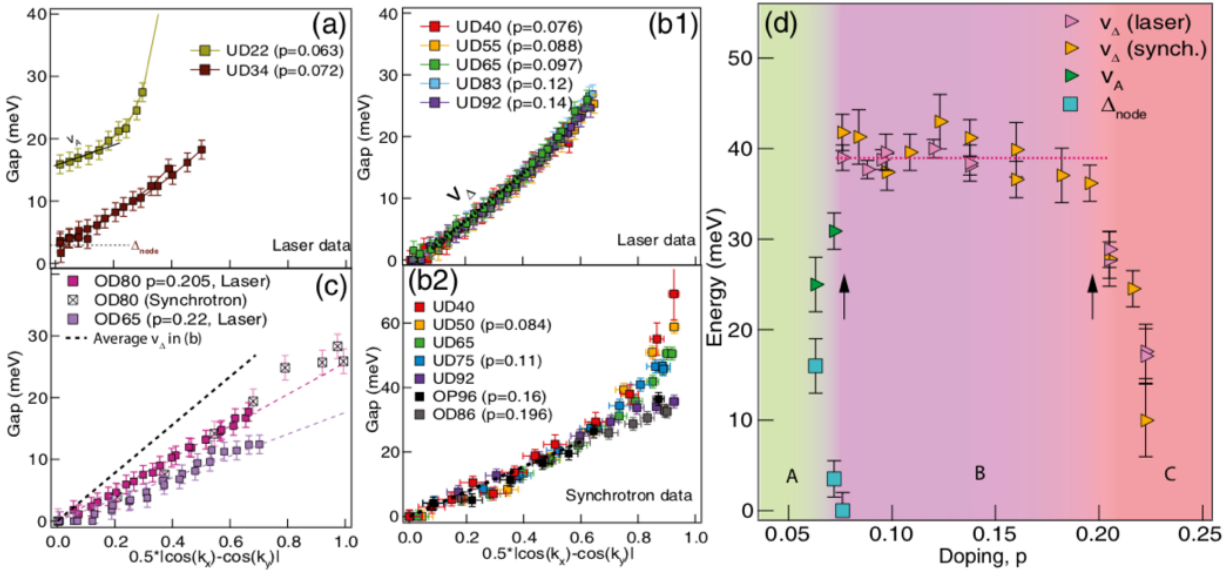

FIG. 1. These ARPES gap spectra for cuprates with the Bi2212 structure duplicate Fig. 2 of [2], where their p comes from $T_c$ using $T_c = 96\text{ K }(1-82.6\,(p-0.16)^2)$. (a-c) The gaps peaks taken at 10 K extrapolate to their largest values at the AN momenta, i.e., along lobes of the d-wave orbitals, for which $K = 0.5|\cos(k_x)-\cos(k_y)| = 1$. Gap nodes appear at the nodal momenta $K = 0$ at 10 K, *except* in Region A (a). (d) their analysis of the near-nodal gap data trisect the SC dome versus doping p into Regions A, B and C, where Region C (p > 0.19) is purely SC as it lacks a high-T PG.

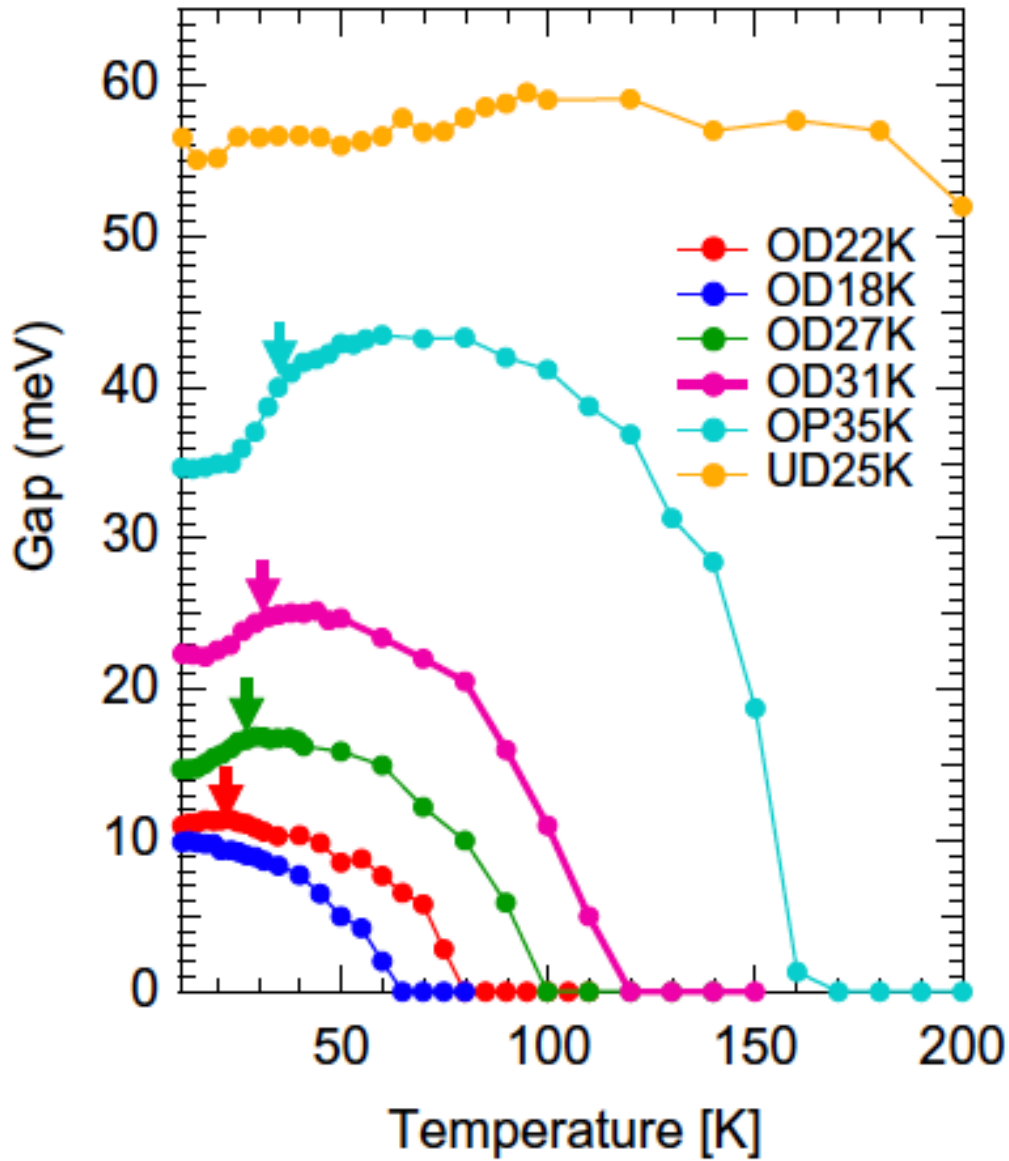

FIG. 2. These AN-ARPES gap peaks for Bi2201 duplicate Fig. 4b of [3]. Note the PG at high-T, which is replaced below $T_c$ (arrows) by a smaller gap that would be consistent with a hybrid gap of the hidden phase of Region B. The bottom curve may be in Region C of Bi2201 which would lack a PG.

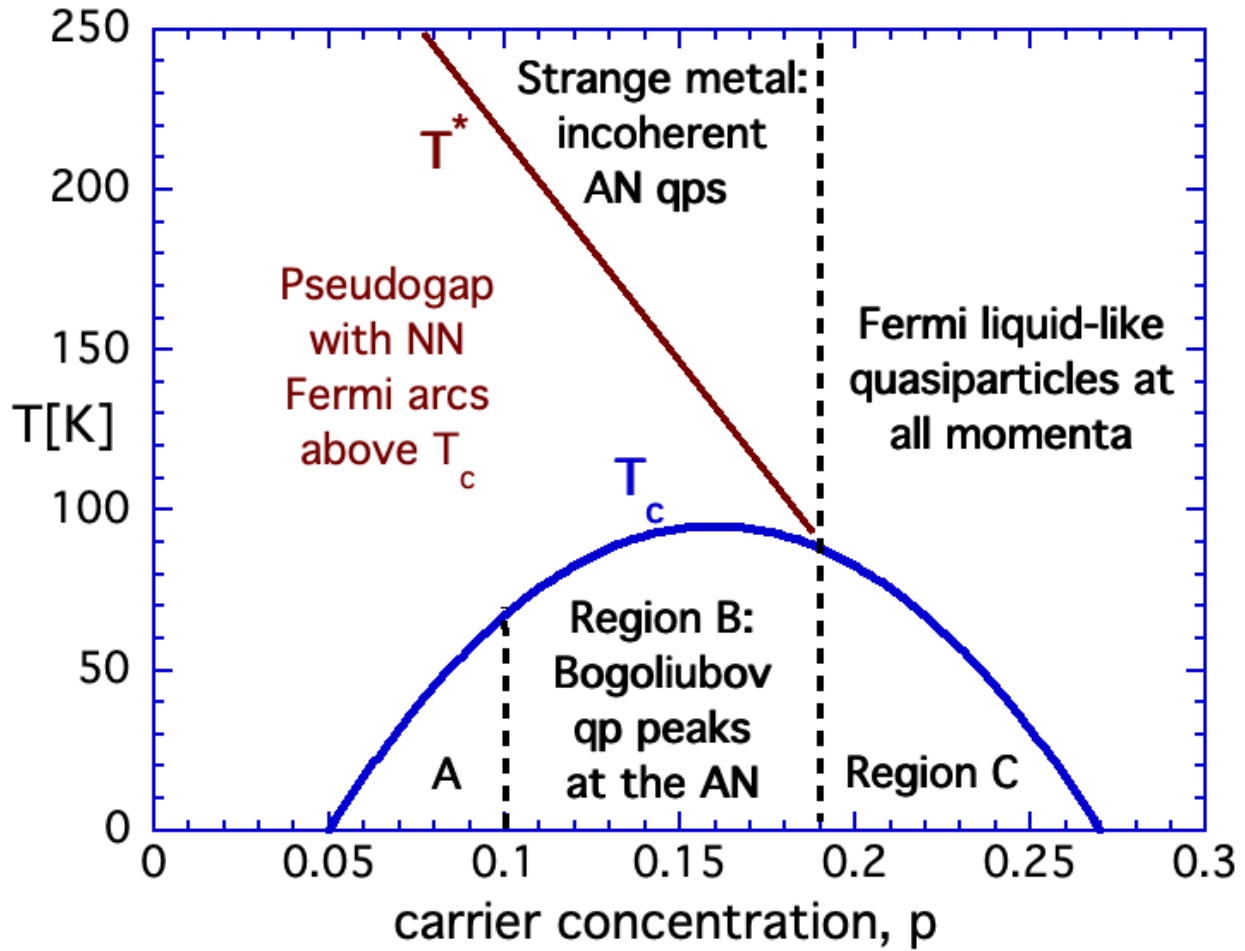

FIG. 3. This figure combines conclusions from various ARPES studies of Bi2212. Notably, the SC dome is trisected into Regions A, B and C [2,3]. For Region B, NN Fermi arcs and a PG are seen for T between $T_c$ and $T^*$ [2,6]; while a strange metal with incoherent AN quasiparticles is seen in the AN-ARPES data at 240 K [4] for p = 0.16 to 0.186. For p = 0.186, coherent Bogoliubov quasiparticle peaks and a dip-hump structure begin to emerge from the high-T strange metal [4] around 110 K and they sharpen considerably below $T_c$ = 86 K. The PG disappears for p > 0.19, and Fermi liquid behavior is seen at all T [4] to confirm that SC condenses onto a Fermi-liquid normal-phase in Region C.

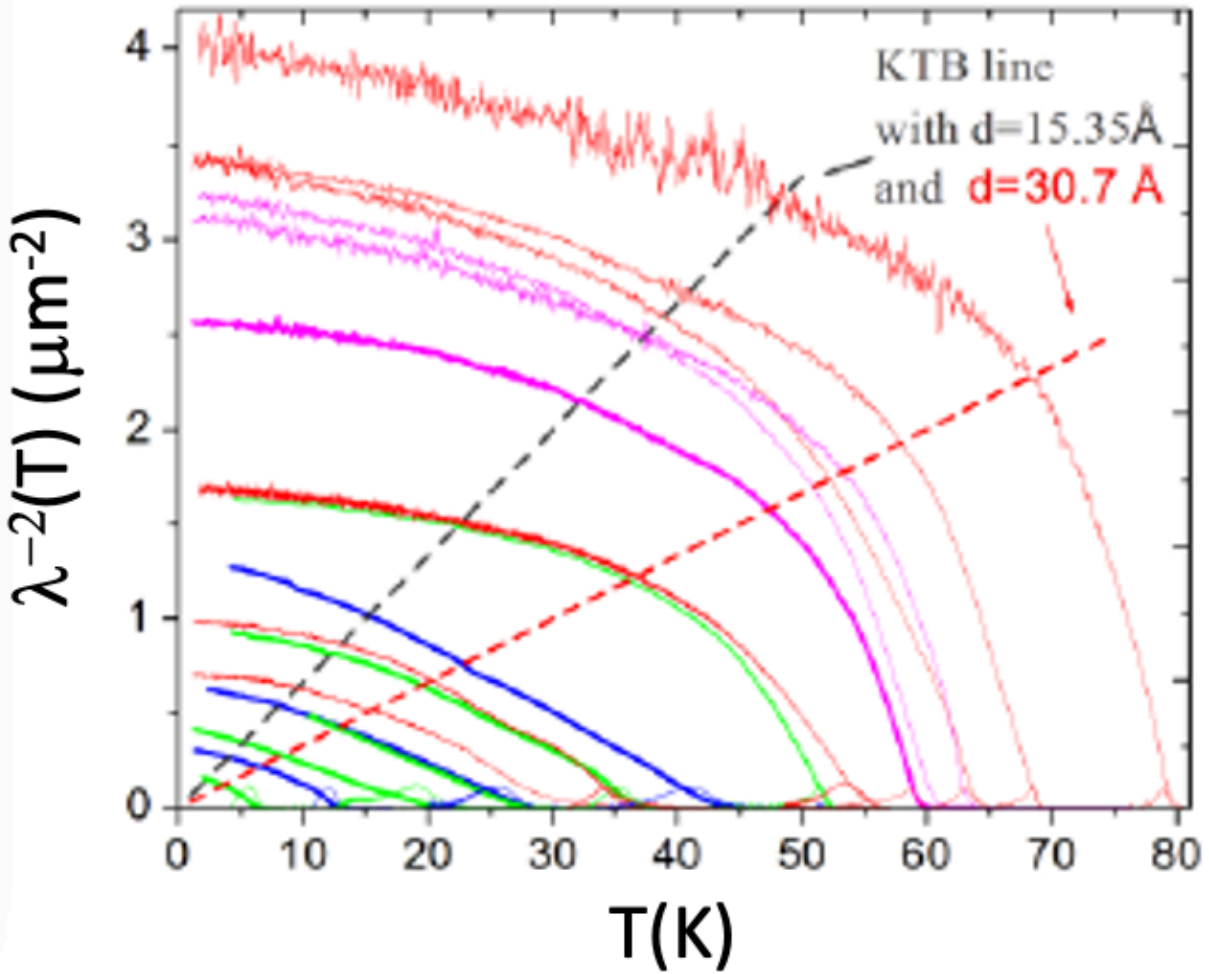

FIG. 4. The superfluid density $n_s(T) = (mc^2/4\pi e^2)/\lambda(T)^2$ is quantified, in units of $\mu m^{-2}$, by the inverse square of the magnetic field penetration depth $\lambda(T)$. The SFD data in this figure are for severely underdoped Bi2212 films cover a doping range $0.055 \leq p \leq 0.12$ and is directly copied from [8]. The downturns for $T_c > 50$ K were first ascribed in [8] to 2D melting theory [9] (dashed KTB lines), but the authors' later analysis [10] seemed to downplay this, and Appendix C rejects it.

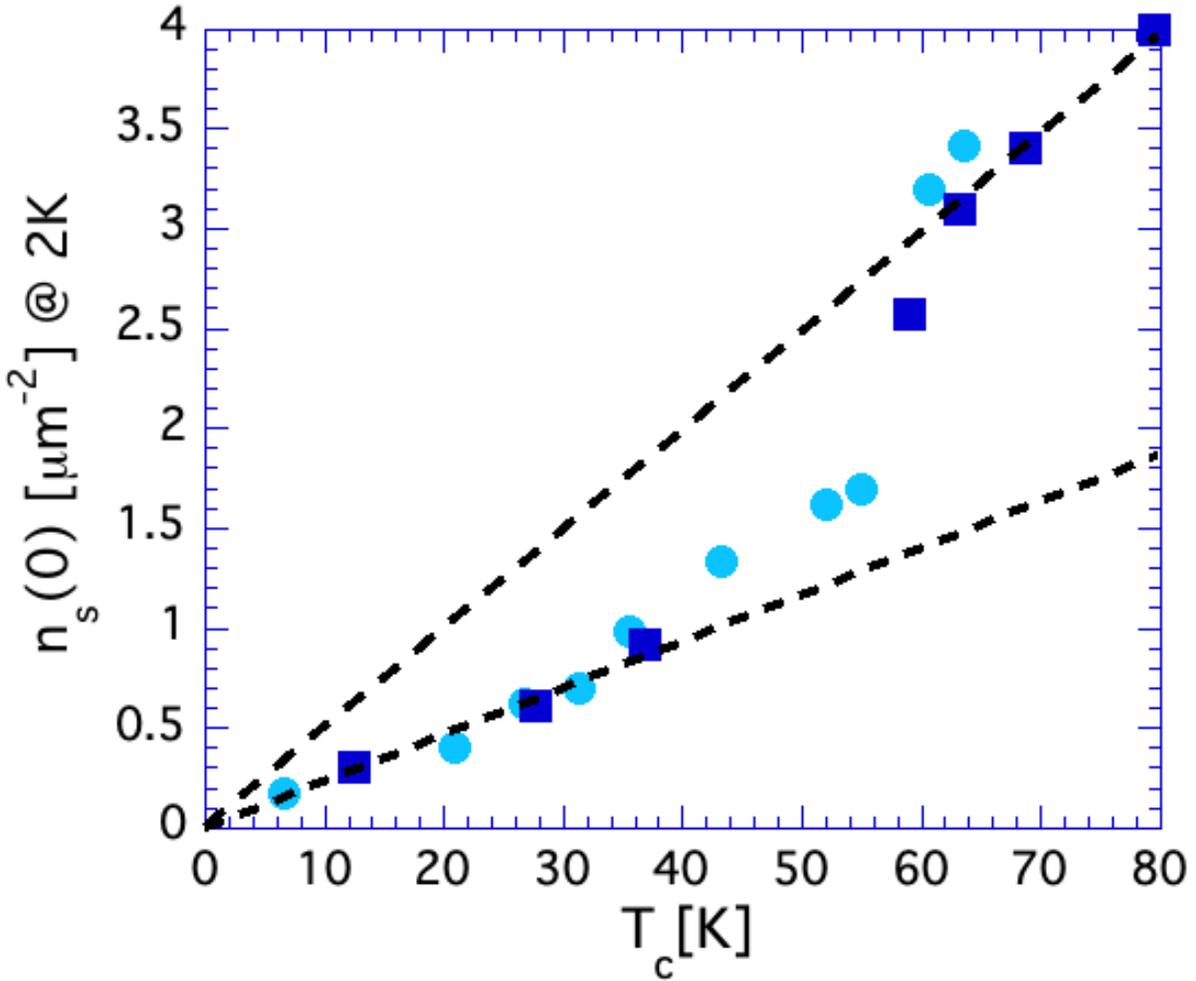

FIG. 5. The low-T $n_s(0)$ from Fig. 4 reveal $n_s(0)$ is proportional to $T_c$, at both small and large $T_c$. Squares denote the most systematic 'parallel curve' data which also have dashed-line fits in Fig. 9. Note that the ratio $n_s(0)/T_c$ in Region B ($T_c > 50$ K) with the hidden phase is twice that of Region A ($T_c < 50$ K), which lacks the hidden phase. These data imply that the p-threshold for doubling of $n_s(0)/T_c$ might be at $T_c \approx 55$ K, or $p \approx 0.09$. Inevitable doping disorder may smear out the 50 K threshold; those data sets in Fig. 4 that are parallel, like curved train tracks, are shown as squares.

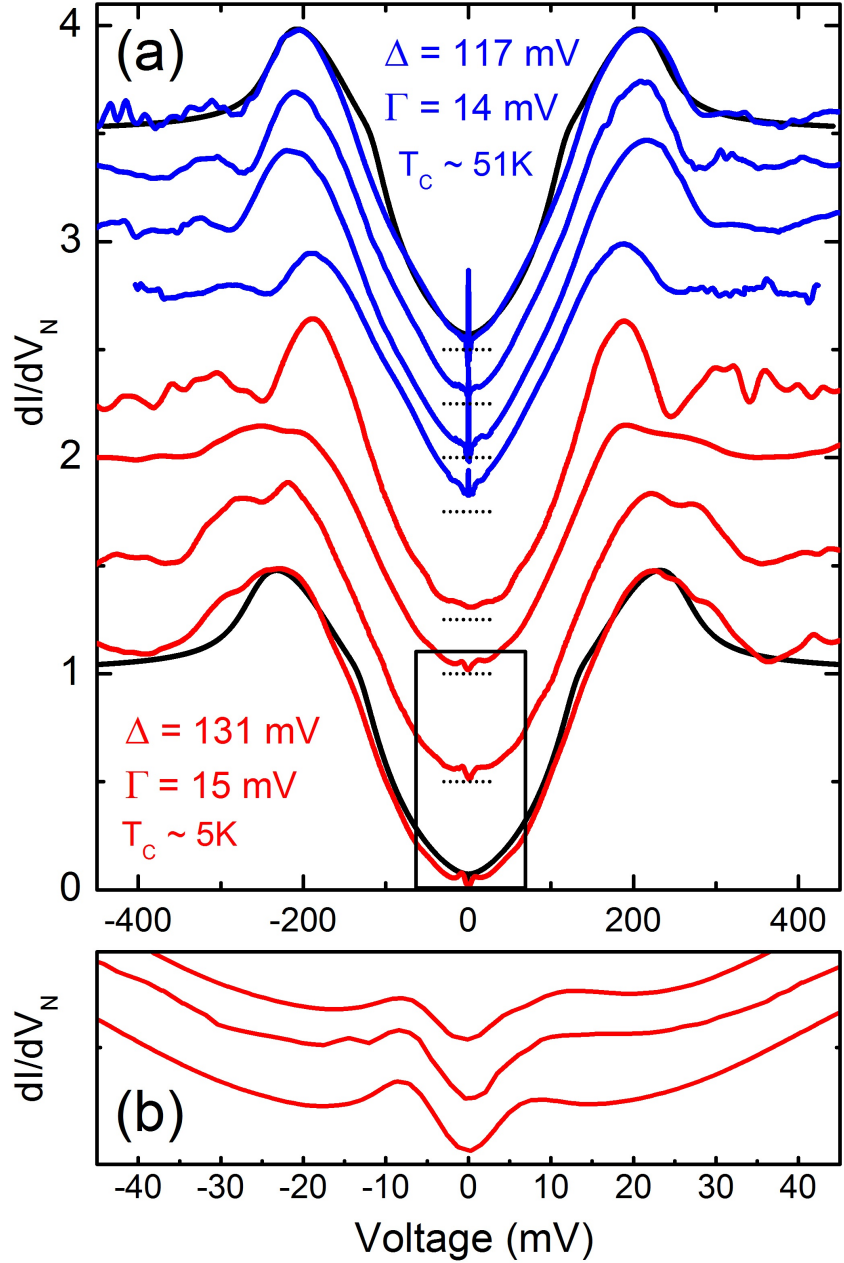

FIG. 6. (a) SIS tunneling conductance data for two severely underdoped Bi2212 films are directly copied from [14]. The upper four curves represent four junctions displaying the largest peaks at $V = \pm 2\Delta/e$ for a film with $T_c \approx 51$ K and $p \approx 0.085$, while the spikes at zero bias result from Josephson supercurrents. The lower four curves are junctions on a film with $T_c \approx 5$ K and $p \approx 0.053$. They are taken well below the nominal $T_c$ at 2 K. All dotted horizonal lines near V = 0 indicate the zero of conductance that would be expected to be found a finite hard gap. (b) The data for $p \approx 0.053$ are magnified to more clearly show the *additional*, much smaller gap peaks.

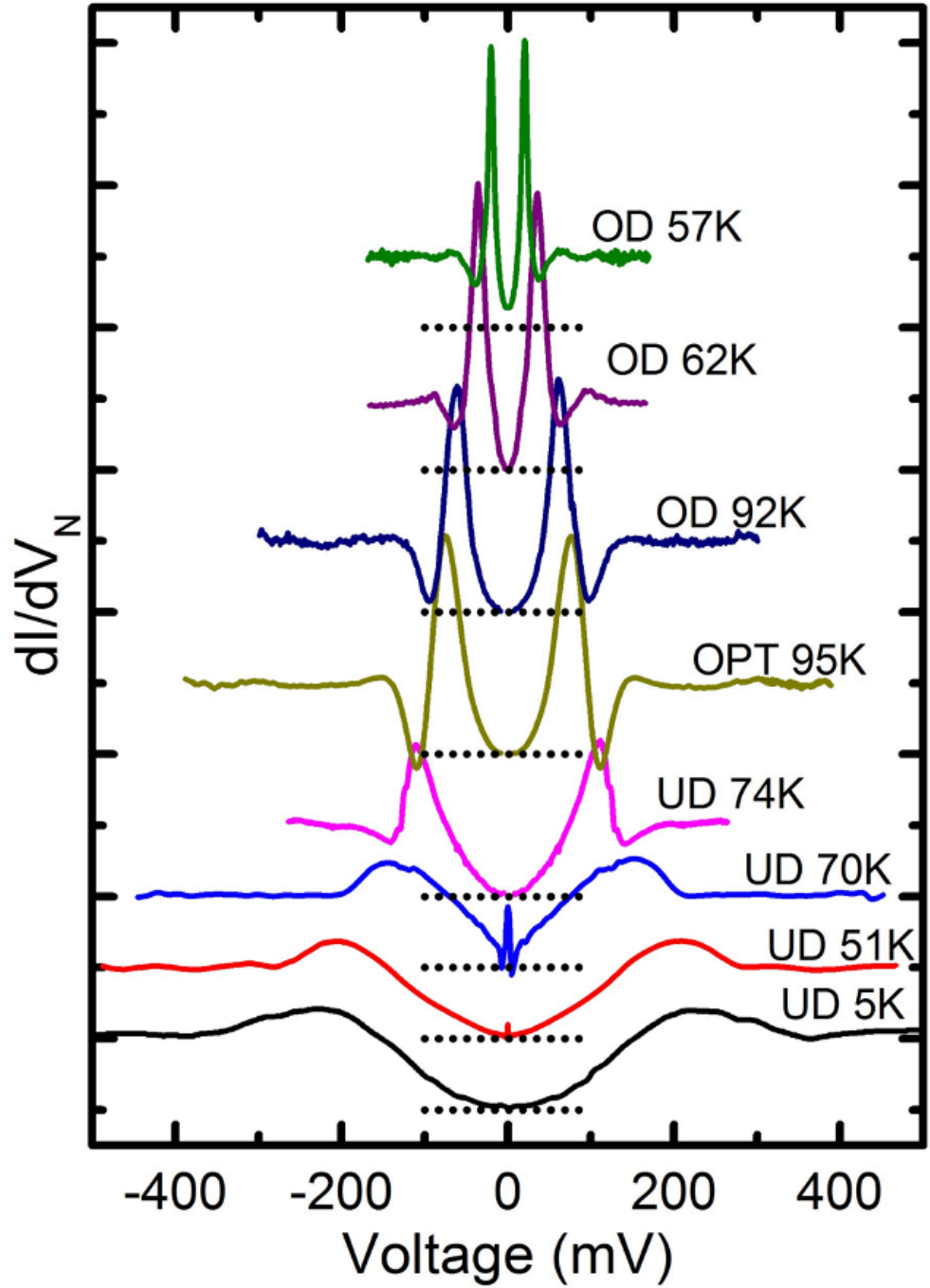

FIG. 7. Tunneling conductances for a broad range of p from [7,13,14] show well-defined dip-hump features at $V \approx \pm 3\Delta/e$ [7] that are absent for $p < 0.11$. Labels on curves: OD is over doped, OPT is optimum doping, UD in underdoped, followed by the SC transition temperature, $T_c$.

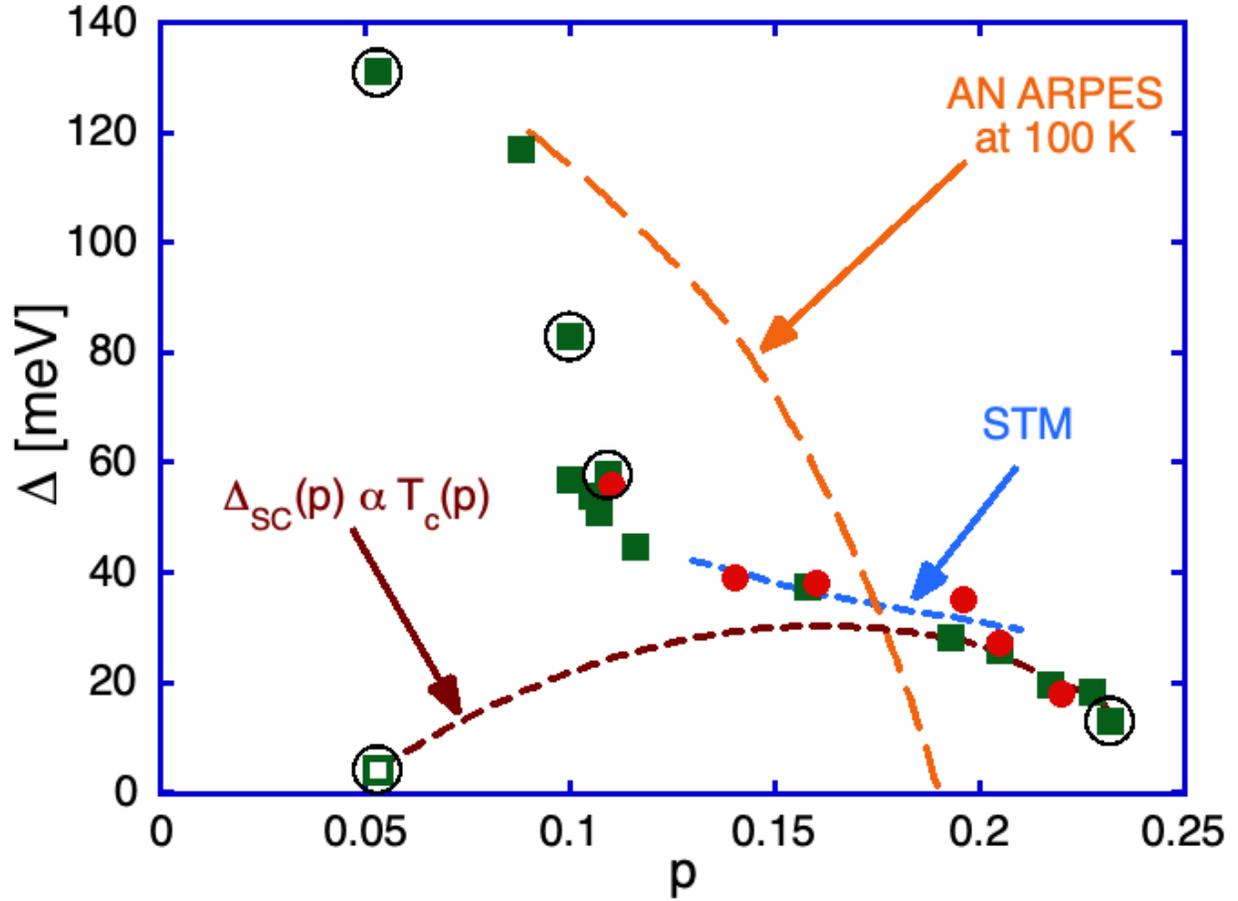

FIG. 8. The k-integrated low-T SIS gap peaks [7,13,14] (squares) for Bi2212 are compared to (i) extrapolated AN-ARPES gap peaks from [2] (circles), (ii) the AN-ARPES gaps at 100 K [6] and (iii) the low-T STM gaps [16]; the latter two, for clarity, are smoothed approximations to those data. Note that two gap peaks are *only* found for $p \approx 0.053$ (see Fig. 6) and the smaller gap appears as an open square. The lowest short-dashed curve is a suggestive fit of $\Delta_{SC}(p)$ proportional to $T_c(p)$ for those data in which SC condenses directly onto a high-T normal state. For the encircled SIS squares, $T_c$ values {and thus p from $T_c = T_{cmax} (1-82.6 (p - 0.16)^2)$ with $T_{cmax}$ = 96 K} results from the high-T disappearance of either the gap structure in Fig. 6b or the Josephson supercurrents, e.g., spikes in Fig. 6a. As such, the encircled squares establish both $\Delta$ and p in the identical microscopic junction volume. Limits on the precision of p-values for all other data are discussed in Appendix D.

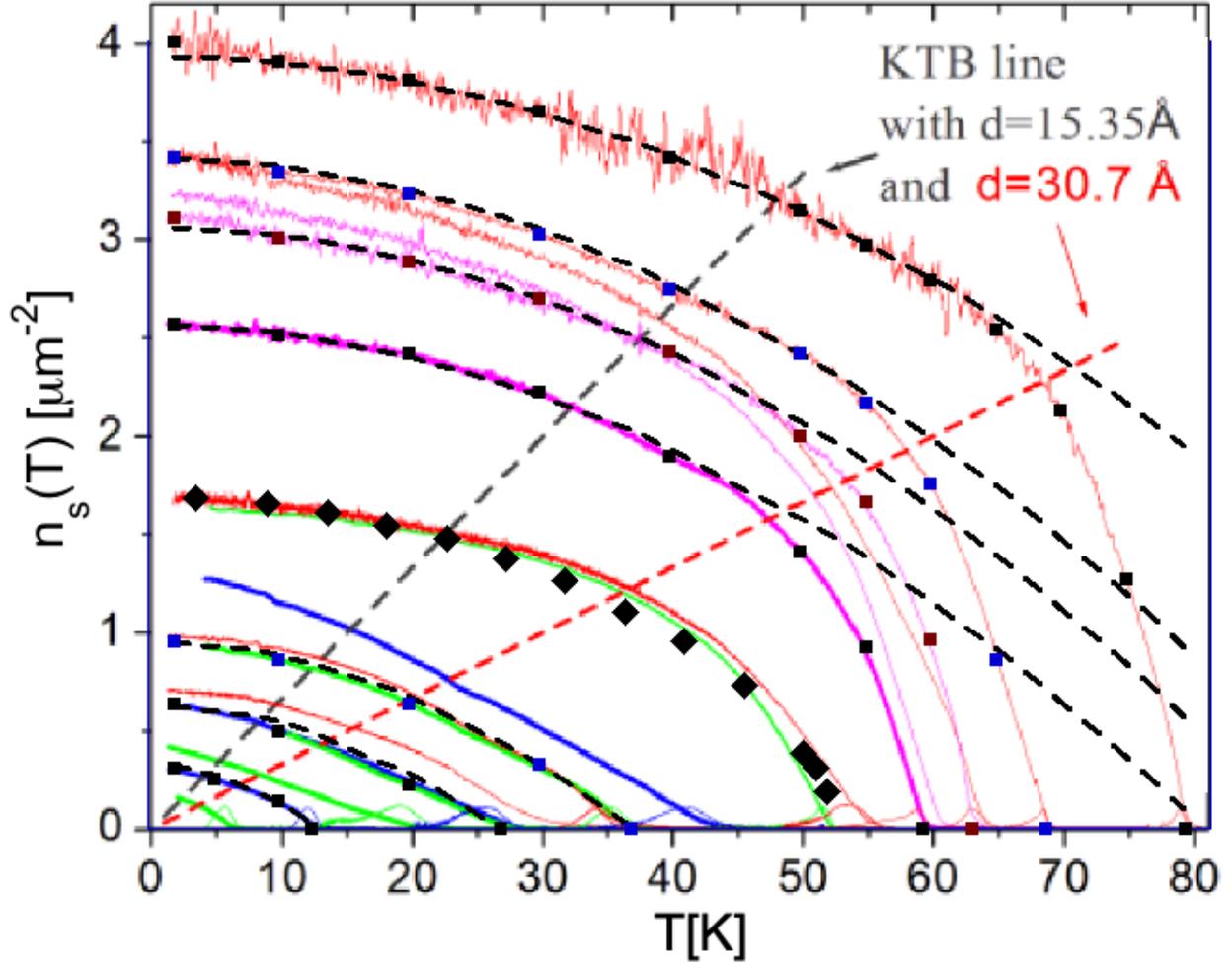

FIG. 9. We added the dashed lines to Fig. 4, from [8], to parameterize $n_s(T)$ by $n_s(0)(1-(T/T_{c2})^2)$; these use fits at low T for data with downturns and at all T for data without a downturn. For the upper four dashed lines, $T_{c2}$ = 81, 89, 97 and 112 K. The diamonds are a fit to the $T_c \approx 54$ K data using the mixed-phase, doping-disorder model described in Sec. 3.4.

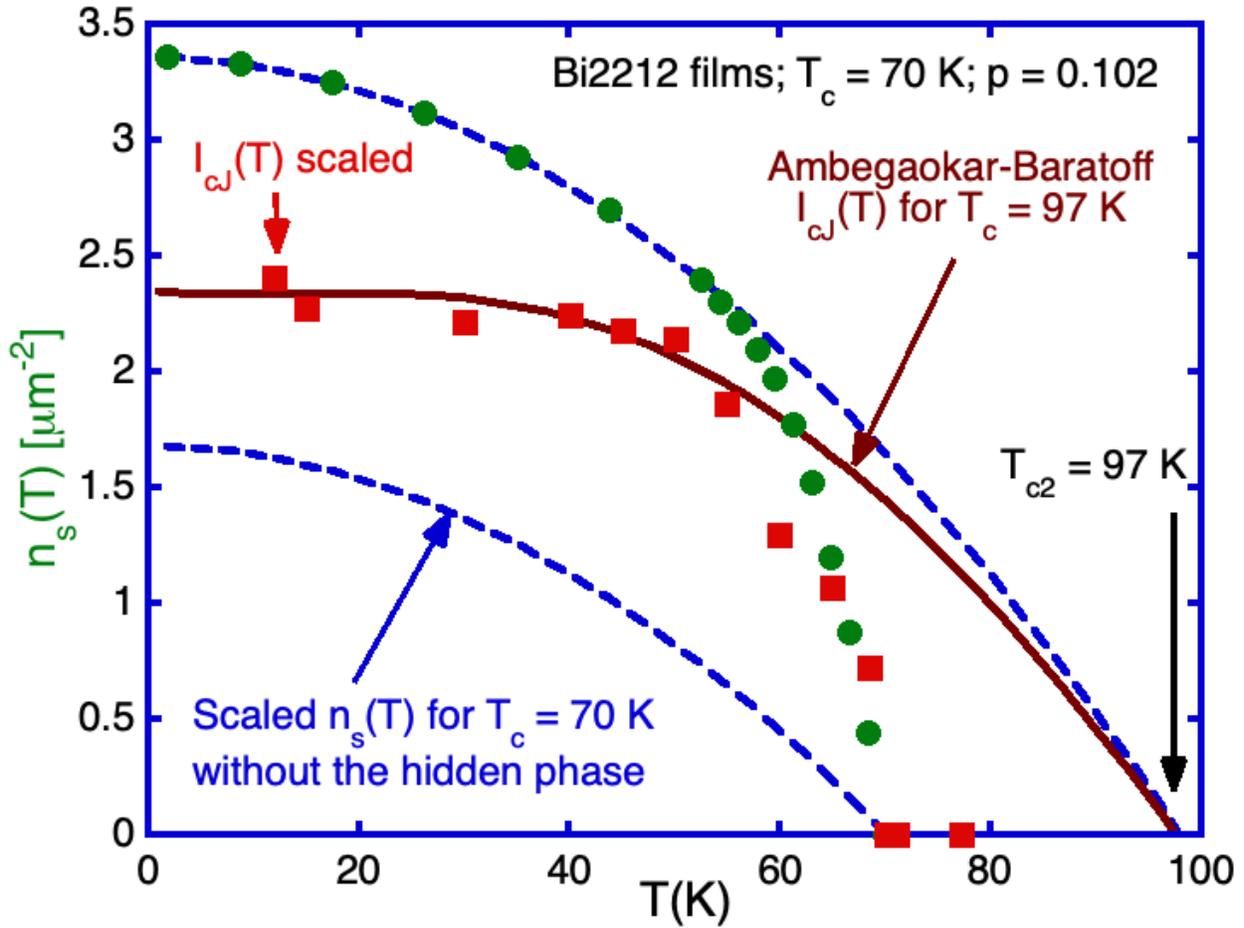

FIG. 10. For films with $T_c \approx 70$ K, the circles denote $n_s(T)$ from [8], while the parametrized dashed curve simulates the novel phase filling the entire film area up to $T_{c2}$. Values of $n_s(T)$ appropriate to $T_{c2}$ are recovered at about 10 K below their threshold at $T_c = 70$ K. That difference implies p-disorder of ± 0.01 around the nominal value of 0.102 for this film which percolates SC at $T_c$. The squares are Josephson supercurrents, $I_{cJ}$, in another film [14] with $T_c \approx 70$ K. These $I_{cJ}$ are scaled for clarity and are seen to be a *proxy* for the SFD as the downturns and the transition widths of each match. Also, the low-T $I_{cJ}$ are consistent with the Ambegaokar-Baratoff theory [19] using the same $T_{c2}$ of 97 K. The lower dashed curve is the expectation for $T_c = 70$ K, *without the hidden phase*, and $n_s(0)$ is half as large, as suggested by the dashed lines in Fig. 5.

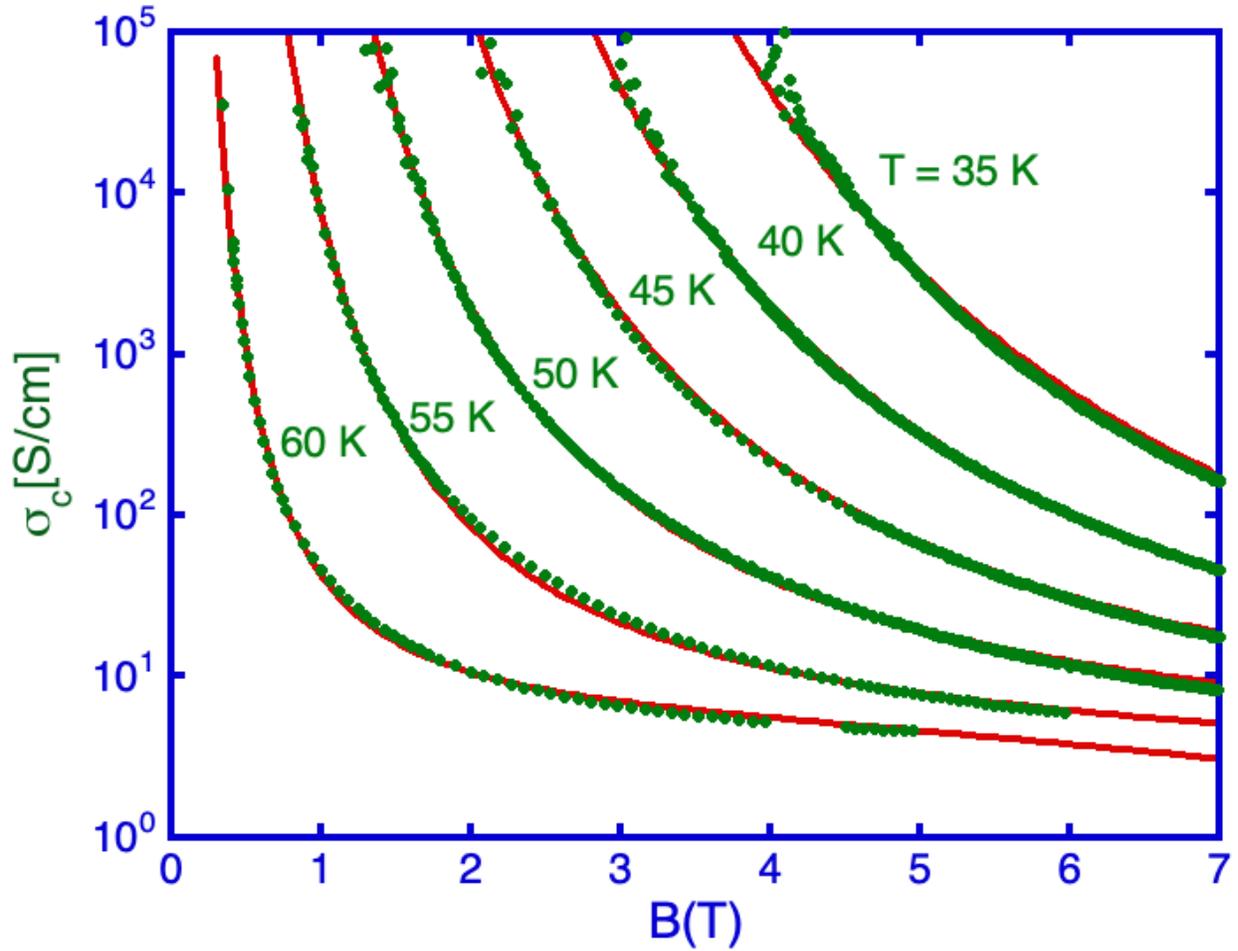

FIG. E.1. The c-axis conductivity (dots) of an underdoped Y123 crystal ($T_c$ = 65 K, $p \approx 0.1$), as a function of T and c-axis magnetic field B, come from [17]. The lines are fits using Eq. E.2 which extends the Ambegaokar-Halperin theory [18] to include the effect of vortex cores at finite B.

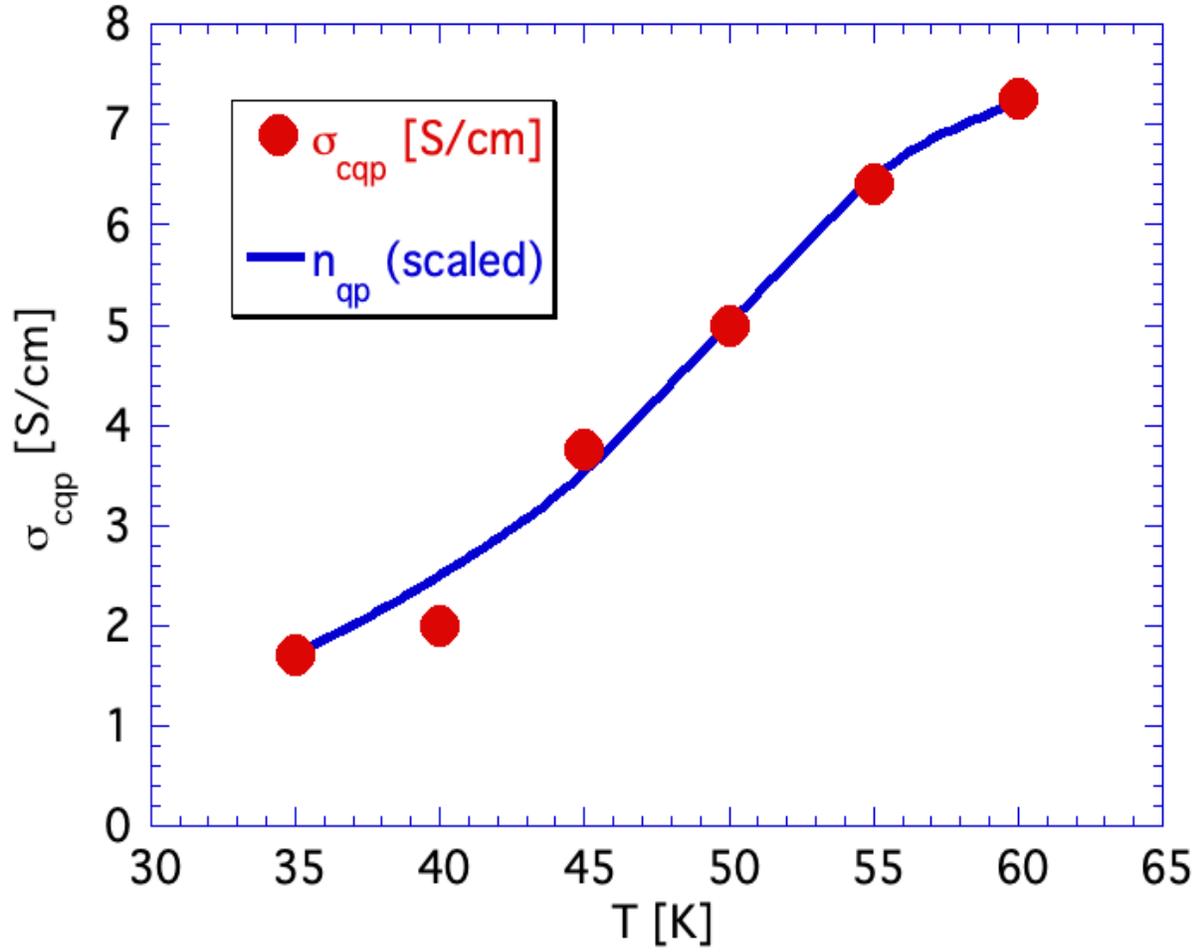

FIG. E.2. Circles are the c-axis quasiparticle conductivities, $\sigma_{cqp}$, that are fitting parameters for our fits of Fig. E.1 to the data of [17] using Eq. E.2. The line is the number density of quasiparticles, $N_{qp}(T)$, calculated as the integral of the Fermi function times the reduced BCS density-of-states summed over the Fermi surface with a d-wave energy gap. These are clearly proportional.